\documentclass[review]{elsarticle}

\usepackage{hyperref}
\usepackage{pbox}
\usepackage{graphicx}
\usepackage{amsmath}
\usepackage{balance}
\usepackage[byname]{smartref}
\usepackage{amsmath,amssymb,amsfonts} 
\usepackage{color}                    
\usepackage{comment}
\usepackage{mathtools}
\usepackage{commath}
\usepackage{graphics,graphicx}
\usepackage{epstopdf}
\usepackage[font={small,it}]{caption}
\usepackage{placeins}
\usepackage{tikz}
\usepackage{bchart}
\usepackage{enumitem}
\usepackage{adjustbox}
\usepackage{pstricks,pst-node,pst-tree}
\usepackage{pst-plot}
\usetikzlibrary{calc,trees,positioning,arrows,chains,shapes.geometric,%
	decorations.pathreplacing,decorations.pathmorphing,shapes,%
	matrix,shapes.symbols,shadows,fit}
\pgfdeclarelayer{background}
\pgfdeclarelayer{foreground}
\pgfsetlayers{background,main,foreground}

\usepackage{pgfplots, pgfplotstable}

\pgfplotsset{compat=1.14, enlarge x limits={abs=0.5},} 
\usepgfplotslibrary{
	units, 
	groupplots
}

\usepackage{tkz-fct}
\usetkzobj{all}

\journal{Knowledge-Based Systems}

\begin{document}

\begin{frontmatter}

\title{Systematic Ensemble Model Selection Approach for Educational Data Mining}


\author[UWO]{MohammadNoor Injadat}
\ead{minjadat@uwo.ca}

\author[UWO]{Abdallah Moubayed}
\ead{amoubaye@uwo.ca}

\author[shj,UWO]{Ali Bou Nassif\corref{mycorrespondingauthor}}
\cortext[mycorrespondingauthor]{Corresponding author}
\ead{anassif@sharjah.ac.ae}

\author[UWO]{Abdallah Shami}
\ead{abdallah.shami@uwo.ca}

\address[UWO]{Electrical \& Computer Engineering Dept., University of Western Ontario, London, ON, Canada}
\address[shj]{Computer Engineering Dept., University of Sharjah, Sharjah, UAE}

\begin{abstract}
A plethora of research has been done in the past focusing on predicting student’s performance in order to support their development. Many institutions are focused on improving the performance and the education quality; and this can be achieved by utilizing data mining techniques to analyze and predict students' performance and to determine possible factors that may affect their final marks. To address this issue, this work starts by thoroughly exploring and analyzing two different datasets at two separate stages of course delivery (20\% and 50\% respectively) using multiple graphical, statistical, and quantitative techniques. The feature analysis provides insights into the nature of the different features considered and helps in the choice of the machine learning algorithms and their parameters. Furthermore, this work proposes a systematic approach based on Gini index and p-value to select a suitable ensemble learner from a combination of six potential machine learning algorithms. Experimental results show that the proposed ensemble models achieve high accuracy and low false positive rate at all stages for both datasets.
\end{abstract}


\end{frontmatter}


\section {Introduction}
The advancement of technology and the Internet has significantly affected learning and education. Within that context, e-learning was developed and can be defined as “the use of computer network technology, primarily over an intranet or through the Internet, to deliver information and instruction to individuals” \cite{00}\cite{7}. However, there are various challenges regarding e-learning, such as the assorted styles of learning, and challenges arising from cultural differences \cite{9}. Other challenges include pedagogical e-learning, technological and technical training, and the management of time \cite{10}. That is why the need for more personalized learning has emerged. 

Personalized learning can be considered as one of the biggest challenges of this century \cite{11}, where the personalization of e-learning includes adaptation of courses to different individuals. One of the biggest learning differences includes the level of knowledge an individual has, and it is being assessed through the learner profile. Learner profile is the most crucial step of the personalization process \cite{KBS3}\cite{11}. To make learning more personalized, adaptive techniques can also be implemented \cite{12}, \cite{13}. Data can be automatically collected from the e-learning environment \cite{13} and then the learner’s profile can be analyzed. 

As part of the learner profile analysis process, predicting student performance plays a crucial role as it can help reduce and prevent student dropout. This is particularly important in an e-learning environment given that it was reported in 2006 that students were 10\% to 20\% more likely to dropout of online courses than traditional classes \cite{s17}. High dropout rates can effect the future of colleges and universities, because policymakers, higher education funding bodies, and educators consider dropout rates to be an objective outcome-based measure of the quality of educational institutions \cite{s18}. Australia, the European Union, the United States of America, and South Africa all use dropout rates as an indicator of the quality of colleges and universities \cite{s19}. As such, universities need to be able to provide accurate analysis of learner profiles and prediction of student performance as well as customize their courses according to the participants' needs \cite{13}, \cite{14}, \cite{15}, \cite{KBS2}. In turn, this can improve the universities' enrollment campaigns and student retention efforts resulting in a higher quality of education \cite{29}.

To analyze collected data, the field of data mining (DM) and machine learning (ML) has emerged and garnered attention in recent years. DM is best defined as an extraction of data from a dataset and discovering useful information from it \cite{1}. Data collected is then analyzed and used for enhancing the decision-making process \cite{2}. DM uses different algorithms in an attempt to establish certain patterns from data \cite{3}. ML and DM techniques have proved to be effective solutions in a variety of fields including education, network security, and business \cite{sc,sd,se}. More specifically, a new sub-field named Educational Data Mining (EDM) has been proposed that focuses on analyzing educational data in order to understand and improve students’ performance \cite{KBS1} and enhance learning and teaching \cite{2}. Data used for EDM includes administrative data, students’ performance data, and activity data \cite{5}. To implement EDM methods, data needs to be collected from different databases and e-learning systems \cite{2}.

Accordingly, this paper uses the comparative analysis gained from various classification algorithms to predict student’s performance at earlier stages of the course delivery. The developed models use ensemble classification techniques to categorize the students and predict their final performance group. In these terms, few classification methods were used, such as K-nearest neighbor (k-NN), random forest (RF), Support Vector machine (SVM), Logistic Regression (LR), Multi-Layer Perceptron (MLP), and Naïve Bayes (NB). These techniques were used individually or as a part of an ensemble learner model to predict the final performance group during the course at two stages - at 20\% and 50\% of the coursework. 
Our research aims to predict the students’ grades during the course as opposed to previous works which focused on performing this analysis at the end of the course. The aim is to identify the best ML individual or ensemble classifier that performs well with e-Learning data. 


The remainder of this work is organized as follows: Section \ref{sec:related_work} describes the related work and its limitations as well as summarizes the research contributions of this work; 
Section \ref{sec:methodology} presents the methodology used for the experiments, highlights and analyzes the utilized datasets, presents the evaluation method, and determines the appropriate parameters for each of the ML algorithms in each dataset; Section \ref{sec:disc} presents and discusses the experimental results; Section \ref{sec:limitation} lists the limitations of the research; and finally, Section \ref{sec:future} concludes the paper and provides some future research direction.

\section{Related Work} \label{sec:related_work}
DM methods have great potential when it comes to analyzing educational data. There is a big interest for understanding the needs of students and their actual level of knowledge. Many researchers have been interested in this problem during the last few years. In 2000, researchers tried to determine low-performing students by using association rules \cite{14}, so that they could involve them in additional courses. Luan \cite{15,16} tried investigating which students are most likely to fail the course by using clustering, neural networks and decision tree methods. In 2003 \cite{17}, Minaeli-Bidgoli et al. used classification for modeling online student grades, while in \cite{18} authors were investigating how students’ performance can be influenced by demographic characteristics and performance.

Pardos et al. \cite{19} used LR to predict the test score in math based on students’ individual characteristics, while Superby et al. \cite{20} used decision tree techniques, RF method, Neural networks and Linear discriminant analysis for predicting students who will most likely drop-out. Vandamme et al. \cite{21} also used Decision tree methods, neural networks and linear discriminant analysis for their prediction of students who will fail the course by classifying them into three groups: low, intermediate and high-risk students. In 2008, Cortez and Silva \cite{22} compared DM algorithms from four different approaches, namely Decision Tree, RF, Neural Network and SVM for prediction of students’ failure.

Kovacic \cite{23} developed a profile of students who would most likely fail or succeed by using classification techniques. He used socio-demographic and learning characteristics as variables for predicting students’ success. Ramaswami et al. \cite{24} tried developing a predictive model that will be used for identifying students who are slow at learning by using Chi-square Automatic Interaction Detector (CHAID) decision tree algorithm. 

Pandey \cite{27} used NB classification to accurately distinguish the bright students from the slow ones. Their model was able to predict students’ grades based on their previous grades. In 2012, authors conducted a comparative research to make a best guess of the student’s performance \cite{28}. The study used decision tree algorithms and it was aimed at finding the best decision tree algorithm that can accurately predict students’ grades. The authors found that CART algorithm that was designed as a decision tree algorithm was the most efficient as it produced the most desired results and concluded that it is desirable to try different classifiers first and then decide which one to use based on the precision and accuracy it gives. Kabakchieva in \cite{29} used four DM algorithms – OneR Rule Learner, Decision Tree, Neural Network and k-NN. Results indicated that the highest accuracy was achieved using the Neural Network algorithm, where the most influencing factors on the classification process were students’ score upon admission and the frequency of failures in the first-year examinations. 

Yadav et al. \cite{30} investigated how the marks from previous or first year exams impact the final grade of engineering students. In their experiments, the authors used classification algorithms such as ID3, J48 (C4.5) and CART and they found that J48 (C4.5) gives the most accurate results. In 2013, one research of secondary education data \cite{31}, performed by using NB and decision tree algorithms, concluded that decision tree classification algorithm was the best for predicting students’ performance and that students’ previous data can be used to predict their final grade.

Hung et al. \cite{32} proposed the use of different classification algorithms such as SVM, RF, and neural networks to improve at-risk student identification. Experimental results performed on two datasets collected from both a school and university environments showed that the proposed approach had a higher accuracy and sensitivity than other works in the literature.

Similarly, Moubayed et al. \cite{ch5ref8a}\cite{35} investigated the problem of identifying the student engagement level using K-means algorithm. Moreover, the authors derived a set of rulers that related student engagement with academic performance using Apriori association rules algorithm. Experimental results analysis showed that the students' engagement level and their academic performance have a positive correlation in an e-learning environment.

Helal \textit{et al.} proposed different classification algorithms to predict student performance while taking into consideration multiple features including socio-demographic features, university admission basis, and attendance type \cite{KBS1}. The authors' experimental results showed that rule-based algorithms as well as tree-based algorithms provided the highest interpretability which made them more useful in an educational environment \cite{KBS1}.

Zupanc and Bosnic extended an existing automated essay evaluation system by considering semantic coherence and consistency features \cite{KBS4}. Through their experimentation, the authors showed that their proposed system provided better semantic feedback to the writer. Moreover, it achieved higher grading accuracy when compared to other state-of-the-art automated essay evaluation systems \cite{KBS4}.

Xu \textit{et al.} proposed a two-layered machine learning model to track and predict student performance in degree programs \cite{ch4_rev1a}. Their simulation results showed that the proposed approach achieved superior performance to benchmark approaches \cite{ch4_rev1a}.

Sekeroglu \textit{et al.} compare the performance of five machine learning classification models to predict the performance of students in higher education \cite{ch4_rev1b}. Their experimental results showed that the prediction performance can be improved by applying data pre-processing mechanisms \cite{ch4_rev1b}.

Khan \textit{et al.} compared the performance of eleven machine learning models in terms of accuracy, F-measure, and true positive rate \cite{ch4_rev1c}. Their experimental results showed that decision tree algorithm outperformed other classifiers in terms of the aforementioned metrics \cite{ch4_rev1c}.
\subsection{Limitations of Related Work}
The difference in the reported results of the previous research is due to multiple factors. First, the participants of the research in different models influence the decision of the studies and their preference. Different researchers have varying interpretation
of the models. Moreover, researchers could be biased depending on the educational environment under consideration. Contradicting results could also be caused by prior knowledge of the researchers concerning the models. To carry out a research, one goes through literature from past studies and in doing so, their stand on the best model could have been biased. Also, the difference in results in related work is because they are not using the same dataset or the same sample in the case where the dataset is the same. The same models perform differently when evaluated using different datasets. Moreover, one major limitation that many of the previous works in the literature suffer from is the fact that they use data collected from one course/term to predict the performance of students in future courses/ terms. However, to the best of our knowledge, none of the previous works predict the student performance during the course delivery.\\ 
After going through the related work, our research aims to confirm the claims and clear any doubts concerning the best model that can identify students who may need help during a course at two stages. By conducting a practical research, our study aims to evaluate the prior findings and their authenticity. Our study will not be biased in any manner and it will look into the nature of datasets. Moreover, our research explores all the six algorithms equally, and any possible ensemble learner that might be developed using these algorithms. The study design predicts the students' grades during the course as opposed to other designs that prefer to conduct it at the end of the course because it is a more accurate predictor. The research assumes that the efforts and seriousness of a student are directly proportional to the final course performance and grade. Therefore, assuming that all factors are constant, the performance of a student can be accurately predicted in the course of the semester.

\subsection{Contribution of Proposed Research}\label{sec:contribution}
Based on the discussion of the related work limitations, the contributions of this work can be summarized as follows:
\begin{itemize}
	\item Analyze the collected datasets and their corresponding features using multiple graphical, statistical, and quantitative techniques (e.g. probability density function, decision boundaries, feature variance, feature weights, principal component analysis, etc.)
	\item Conduct hyper-parameter tuning to optimize the parameters of the different ML algorithms under consideration using grid search algorithm. 
	\item Propose a systemic approach for building an unbiased (through multi-splits) ensemble learner to choose the best model based on multiple performance metrics, namely the Gini index and the p-value.
	\item Evaluate the performance of traditional classification techniques compared to the proposed ensemble learner, and identify students who may need help with high accuracy using the proposed ensemble learner.
\end{itemize}

\begin{figure}[b!]
	\centering
	\includegraphics[scale=1]{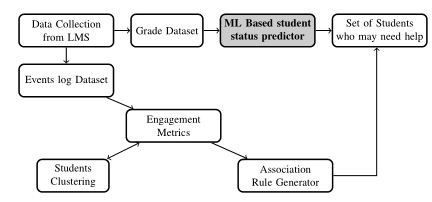}	
	\caption{\label{fig:1.1} Learning Management System (LMS) Analytical Module}
\end{figure}
\section{Methodology and Research Framework} \label{sec:methodology}
\subsection{General Research Framework}
\begin{figure}[!htb]
        \centering
        \includegraphics[scale=0.1]{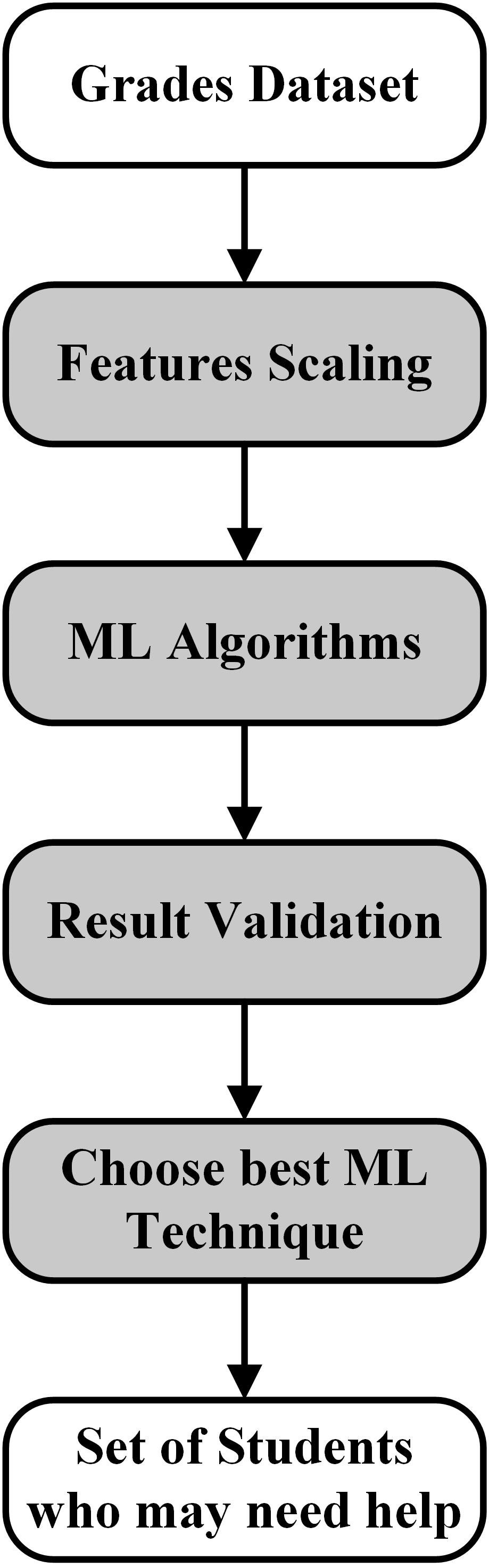}
        \caption{\label{fig:1.2} ML-Based Student Status Predictor}
\end{figure}

The purpose of this study is to predict students’ final grades in order to identify students who may need help at earlier stages of the course. Figure \ref{fig:1.1} shows the analytical process for the data collected through the Learning Management System (LMS). The ``Data Collection from LMS'' module represents the process of collecting data from the LMS. This includes two different types of data, namely the grades of each student (stored in the ``Grade Dataset'' module) and the event log dataset (stored in the ``Events Log Dataset'' module). This work focuses on the student status prediction using ML (highlighted in grey as the ``ML-based student status predictor'' module). More specifically, the  \emph{ML Based student status predictor} is structured as in Figure \ref{fig:1.2}. Note that the ``Engagement metrics'', ``students clustering'', and ``Association rule generator'' modules try to gauge the engagement of students and identify students who may need help. This work was completed in our previous work \cite{35}.

Two datasets were used in this experiment. Dataset 1 consists of records of 52 first year students who completed the undergraduate engineering course (out of 115 registered students) at the University of Genoa \cite{67}. On the other hand, Dataset 2 consists of records of 486 students who attended undergraduate science course at University of Western Ontario, Canada. Event logs to the LMS and students’ individual marks were used in the analysis. Moreover, these experiments predict the final grade based on the individual marks during the course at two stages: $20\%$ and $50\%$ of the coursework.

To improve the accuracy of the prediction, the individual marks were converted to percentage as this scaling of scores (grades) improved the experimental results accuracy. The scaling of scores was also important when it came to compare the performance of students. Furthermore, if a student was absent for a certain mark and it was empty in the dataset, it was replaced with the value of zero. This also improves the experimental results accuracy across the considered techniques.

The final grade was classified into two categories (classes): 
\begin{itemize}
	\item Good (G) – the student will finish the course with a good grade ($60\% -100\%$);
	\item Weak (W) – the student will finish the course with a weak grade ($\leq 59\%$). This limit was chosen since the typical passing grade for undergraduate course is often set to 60 in most universities around the world.
\end{itemize}
The second class represents the targeted learners, i.e. students who need additional assistance and concentration in order to improve their performance. 

\subsection{Datasets Description} \label{sec:dataset}
In the 1950's, an American Educational psychologist, Benjamin Bloom, developed his taxonomy of cognitive objectives. According to \emph{Bloom's Taxonomy}, thinking skills and objectives can be categorized and ordered following the thinking process, \cite{U}. Bloom's Taxonomy was revised years later when the categories or taxonomic elements were associated with it Lower Order Thinking Skills (LOTS):
\begin{itemize}
	\item Remembering - Recognizing, listing, describing, identifying, retrieving, naming, locating, finding 
	\item Understanding - Interpreting, Summarizing, inferring, paraphrasing, classifying, comparing, explaining, exemplifying 
	\item Applying - Implementing, carrying out, using, executing 
	\item Analyzing - Comparing, organizing, deconstructing, Attributing, outlining, finding, structuring, integrating 
	\item Evaluating - Checking, hypothesizing, critiquing, Experimenting, judging, testing, Detecting, Monitoring 
	\item Creating - designing, constructing, planning, producing, inventing, devising, making
\end{itemize}

In this section we describe two datasets at two different stages (20\% and 50\% of the coursework), consisting of the results of a collection of tasks performed by University students, and we conduct some Principal Components Analysis. Interestingly, the first four principal components for Dataset 1, stage 20\% and 50\%, correspond to four of the categories above.\\
\indent In this paper, R was used for numerical analysis, ML techniques, and data visualization \cite{39}. R is a language and environment for statistical computing and graphics.
\begin{itemize}
	\item \emph{Dataset 1}:
	The experiment has been conducted with a group of 115 students of first-year, undergraduate engineering major of the University of Genoa  \cite{67}. The dataset contains data collected using a simulation environment named Deeds (Digital Electronics Education and Design Suite) and it is used in e-Learning courses. The e-Learning platform offers the courses’ contents using a special browser which will ask the students to solve problems that lied under different complexity levels.\\ 
	The records were summarized in Table \ref{tab:table_dataset_1_list} in order to be analyzed with the features' distribution shown in Fig. \ref{fig:data1vars_distrib}. Only 52 students completed the course. \\
	Features ES 1.1 to ES 3.5 were used in the 20\% stage. Features ES 1.1 to ES 5.1 which used at the 50\% stage. 
	Note that the sum of features ES 1.1 to ES 5.1 is 47\% of the course total mark. However, since ES 5.2 had a weight of 10\% and to maintain consistency of performing the analysis at a similar stage during the course delivery among the two datasets, these features were considered at the 50\% stage since their sum is close to that percentage. These features (ES 1.1 to ES 5.1) can be categorized based on their cognitive objectives using Bloom's taxonomy as follows: features ES 1.1 and ES 3.4 belong to the \textit{Understand} category; features ES 2.1, ES 2.2, and ES 3.3 belong to the \textit{Apply} category; features ES 1.2, ES 3.1, ES 3.2, and ES 3.5 belong to \textit{Analyze} category; and finally features ES 2.1, ES 3.4, and ES 3.5 belong to the \textit{Evaluate} category. These features are used to predict the student performance on the remaining tasks/features regardless of their cognitive objective category.\\  
	Any empty mark was replaced with 0. Also, all features were converted to a mark out of 100 which improves the accuracy of all classifiers. Any mark that consist of decimal point number was rounded to the nearest 1.
	\begin{table}[h!]
		\centering
		\caption{Dataset 1 - Features}
		\begin{adjustbox}{max width=\columnwidth}
			\scalebox{0.75}{ 
			\begin{tabular}{|c|c|c|c|} 
				\hline     
				\textbf{Feature} & \textbf{Description} & \textbf{Type} & \textbf{Value/s} \\
				\hline
				\text{Id}           &\text{Student Id.} &\text{Nominal} & \text{Std. 1,..,Std. 52} \\
				\hline
				\text{ES 1.1}    & \text{Exc. 1.1 Mark} &\text{Numeric} & \text{0..2} \\
				\hline
				\text{ES 1.2}    & \text{Exc. 1.2 Mark} &\text{Numeric} & \text{0..3} \\
				\hline
				\text{ES 2.1}    & \text{Exc. 2.1 Mark} &\text{Numeric} & \text{0..2} \\
				\hline
				\text{ES 2.2}     & \text{Exc. 2.2 Mark} &\text{Numeric} & \text{0..3} \\
				\hline
				\text{ES 3.1}     & \text{Exc. 3.1 Mark} &\text{Numeric} & \text{0..1} \\
				\hline
				\text{ES 3.2}     &\text{Exc. 3.2 Mark} &\text{Numeric} & \text{0..2} \\
				\hline
				\text{ES 3.3}        & \text{Exc. 3.3 Mark} &\text{Numeric} & \text{0..2} \\
				\hline
				\text{ES 3.4}        & \text{Exc. 3.4 Mark} &\text{Numeric} & \text{0..2} \\
				\hline
				\text{ES 3.5} &\text{Exc. 3.5 Mark} &\text{Numeric} & \text{0..3} \\
				\hline
				\text{ES 4.1} &\text{Exc. 4.1 Mark} &\text{Numeric} & \text{0..15} \\
				\hline
				\text{ES 4.2} & \text{Exc. 4.2 Mark} &\text{Numeric} & \text{0..10} \\
				\hline
				\text{ES 5.1} & \text{Exc. 5.1 Mark} &\text{Numeric} & \text{0..2} \\
				\hline
				\text{ES 5.2} &\text{Exc. 5.2 Mark} &\text{Numeric} & \text{0..10} \\
				\hline
				\text{ES 5.3} & \text{Exc. 5.3 Mark} &\text{Numeric} & \text{0..3} \\
				\hline
				\text{ES 6.1} & \text{Exc. 6.1 Mark} &\text{Numeric} & \text{0..25} \\
				\hline
				\text{ES 6.2} & \text{Exc. 6.2 Mark} &\text{Numeric} & \text{0..15} \\
				\hline
				\text{Final Grade}&  \text{Total Final Mark} &\text{Numeric} & \text{0..100} \\
				\hline
				\text{Total} &\text{Final Course Grade} &\text{Nominal} & \text{G,W} \\
				\hline
			\end{tabular}
		}
		\end{adjustbox}
		\label{tab:table_dataset_1_list}
	\end{table}
	\begin{figure}[h!]
		\centering
		\includegraphics[scale=0.6]
		{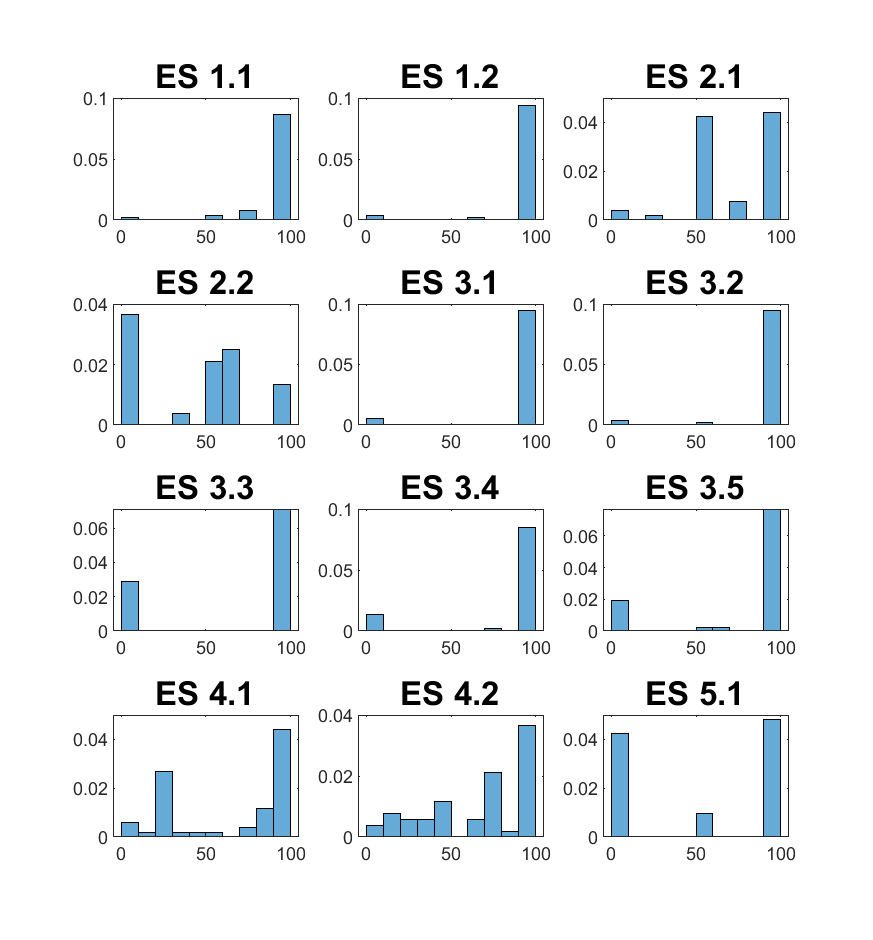}
		\caption{\label{fig:data1vars_distrib} Dataset 1 - Features distribution}
	\end{figure}
	In particular, the Final Grade feature has a distribution as seen in Fig. \ref{fig:fingrdistr20}.
	\begin{figure}[h!]
		\center{\includegraphics[scale=0.35] 
			{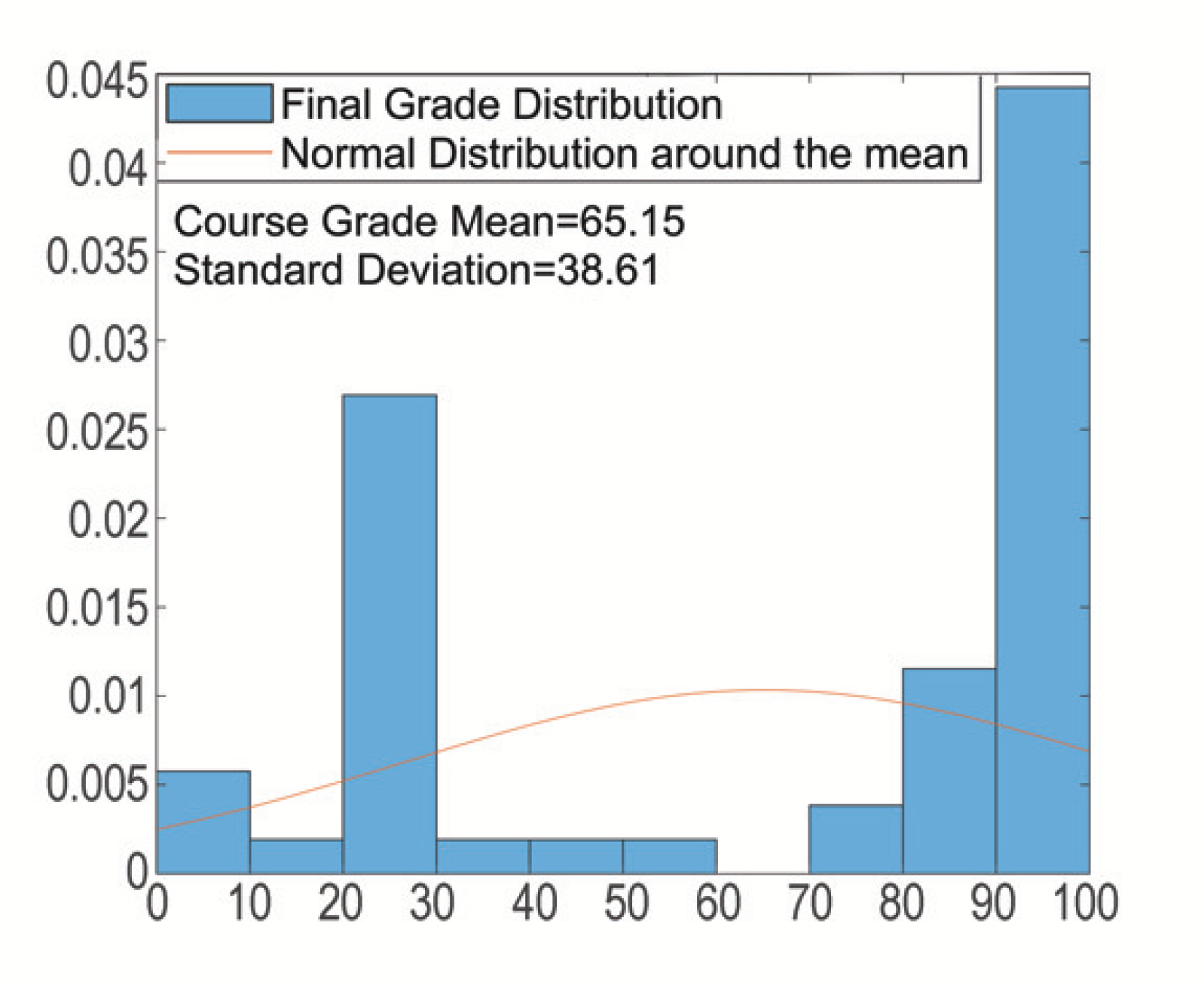}}
		\caption{\label{fig:fingrdistr20} Dataset 1 -  Final Grade Distribution}
	\end{figure}
	\item \emph{Dataset 2}:
	The collected dataset is from a second year undergraduate Science course offered at The University of Western Ontario. The dataset consists of two parts. The first part is an event log of 486 enrolled students and has a total of 305933 records collected from the university’s LMS. The second part which is used in this experiment is the obtained grades of the 486 students in the different assignments, quizzes, and exams. Features Quiz 01 and Assignment 01 were used in the 20\% stage.
	Note that the sum of these two features represents 18\% of the course total mark. However, as mentioned earlier, these features were considered at the 20\% stage since their sum is close to the desired percentage and to maintain consistency of performing the analysis at similar stages during the course delivery among the two datasets. 
	For the 50\% stage, features Quiz 01 to Assignment 02 were used with their sum representing 50\% of the course total mark. Any empty mark was replaced with 0. Also, all features were converted to a mark out of 100 which improves the accuracy of all classifiers. Any mark that consists of decimal point number was rounded to the nearest 1. The total course mark was counted out of 110 with the additional 10\% being added to assignment 03's grade as curving to help students in the course's final grade.  In Table \ref{tab:table_dataset_2}, we show the list of features corresponding to dataset 2.
	Note that due to the recent adoption of the ``\textit{General Data Protection Regulation}'' which introduced many restrictions on the collection of data, and to maintain the privacy of users, the contents of the different tasks/features could not be accessed. As such, these tasks/features could not be categorized as per their cognitive objectives using Bloom's taxonomy.
	\begin{table}[h]
		\centering
		\caption{ Dataset 2 - Features}
		\begin{adjustbox}{max width=\columnwidth}
			\scalebox{0.75}{ 
			\begin{tabular}{|l|p{2.5cm}|c|c|} 
				\hline     
				\textbf{Feature} & \textbf{Description} & \textbf{Type} & \textbf{Value/s} \\
				\hline
				\text{Id}           &\text{Student Id.} &\text{Nominal} & \text{std000,..,std485} \\
				\hline
				\text{Quiz01}    & \text{Quiz1 Mark} &\text{Numeric} & \text{0..10} \\
				\hline
				\text{Assign.01}    & \text{Assign.01 Mark} &\text{Numeric} & \text{0..8} \\
				\hline
				\text{Midterm}    & \text{Midterm Mark} &\text{Numeric} & \text{0..20} \\
				\hline
				\text{Assign.02}     & \text{Assign.02 Mark} &\text{Numeric} & \text{0..12} \\
				\hline
				\text{Assign.03}     & \text{Assign.03 Mark} &\text{Numeric} & \text{0..25} \\
				\hline
				\text{Final Exam} & \text{Final Exam Mark} &\text{Numeric} & \text{0..35} \\
				\hline
				\text{Final Grade}&  \text{Total Final Mark} &\text{Numeric} & \text{0..100} \\
				\hline
				\text{Total} &\text{Final Grade} &\text{Nominal} & \text{G,W} \\
				\hline
			\end{tabular}}
		\end{adjustbox}
		\label{tab:table_dataset_2}
	\end{table}
	
	And the distribution of the features is shown in Fig. \ref{fig:data2vars_distrib}.
	\begin{figure}[h!]
		\centering
		\includegraphics[scale=0.65]
		{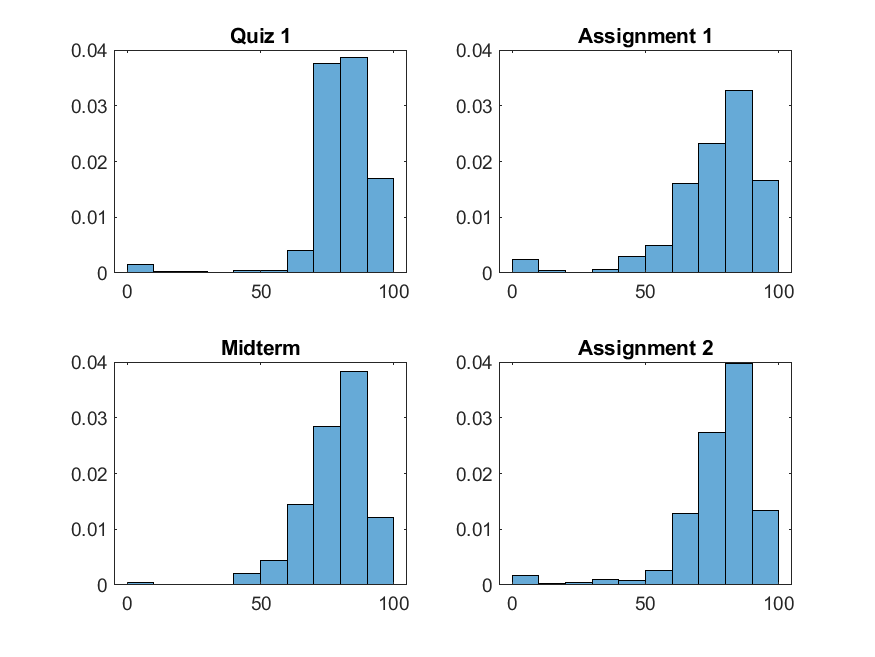}
		\caption{\label{fig:data2vars_distrib} Dataset 2 - Features distribution}
	\end{figure}
	The distribution of the Final Grade of the second dataset is  shown in Fig. \ref{fig:fingrdistr_big}.
	\begin{figure}[!tb]
		\center{\includegraphics[scale=0.38]
			{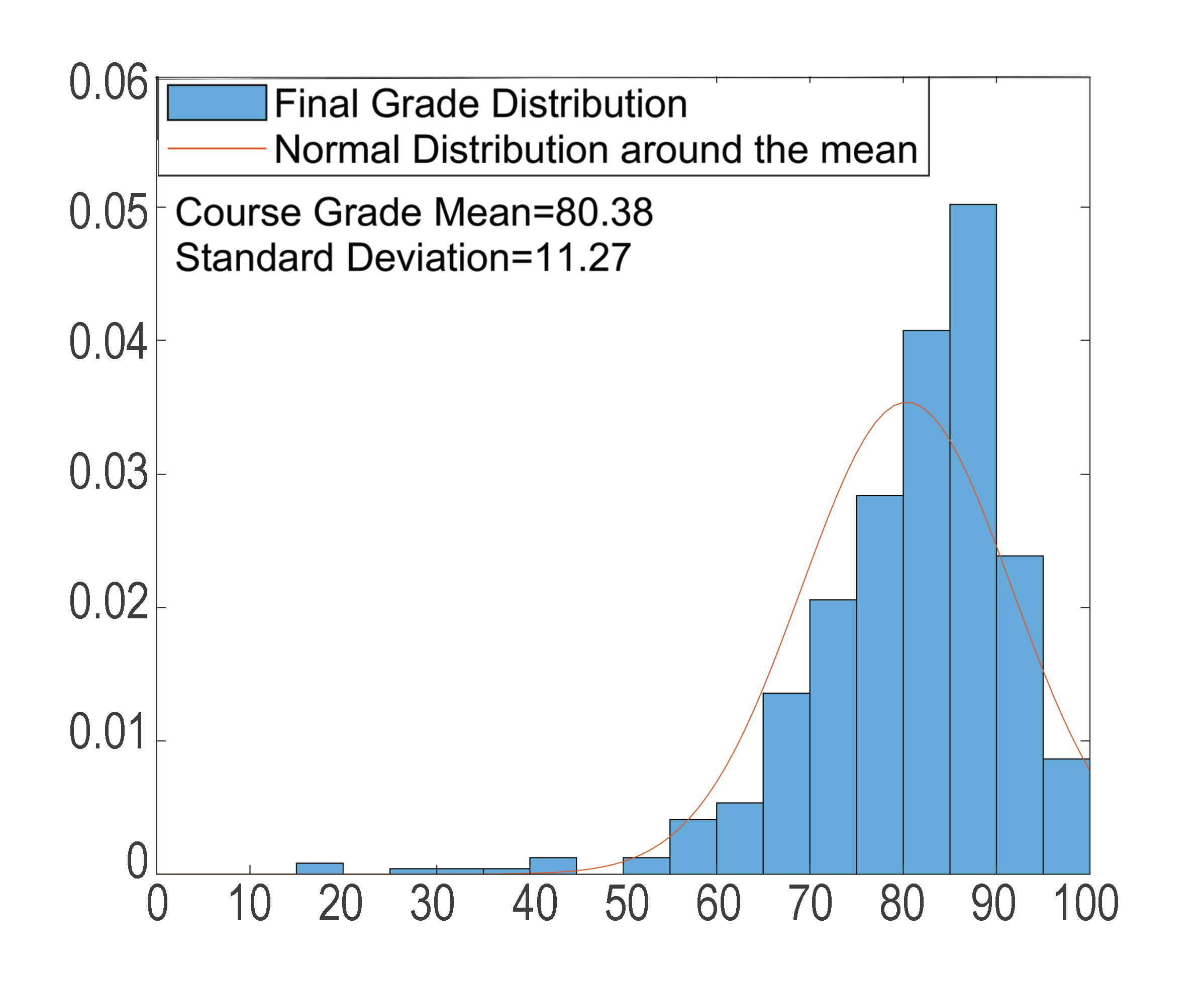}}
		\caption{\label{fig:fingrdistr_big}Dataset 2 - Final Grade Distribution}
	\end{figure}
	\begin{figure}[t!]
		\centering
		\includegraphics[trim=1cm 0cm 0cm 0cm, scale=0.5]
		{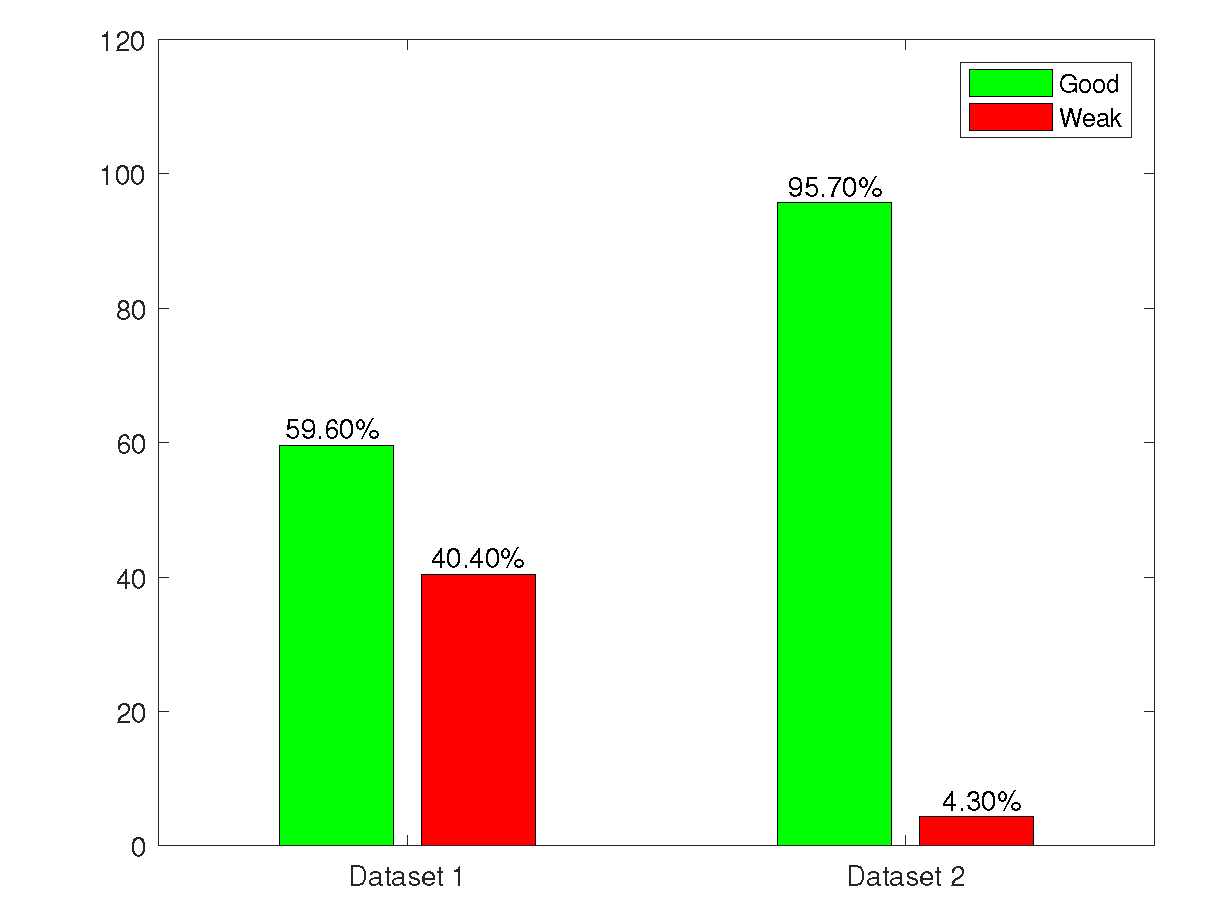}
		\caption{\label{fig:target}Target variable: Dataset 1 vs. Dataset 2}
	\end{figure}
	
	Fig. \ref{fig:fingrdistr20} shows that the first dataset is not normally distributed (due to the fact that only 52 students out of the 115 that were initially registered for the course completed it, thus directly impacting the normal distribution of the final grades) while Fig. \ref{fig:fingrdistr_big} shows that the second dataset has a skewed normal distribution. This means that some classifiers are unlikely to have a good performance on the given datasets. For example, NB, which is a technique that performs very well in case of normally distributed numerical input (not categorical), is not expected to perform well with the considered datasets.
\end{itemize}
Note that the Dataset 2 is unbalanced (4.3\% of Weak students) whereas Dataset 1 has 40.4\% of Weak students, as summarized in Fig. \ref{fig:target}. Note that to overcome the issue of data being imbalanced, multiple procedures were considered in this work. The first is using the Gini index and p-value as evaluation metrics as this makes the reported results more robust and statistically significant. The second is using multiple splits to reduce the bias in the obtained results. Last but not least, the performance was evaluated using the specificity and sensitivity rather than the accuracy since these metrics better illustrate the performance of the classifiers when dealing with imbalanced data.
\subsection{Dataset Visualization}\label{Dataset_Visualization}
In ML problems, it is very important to visualize the dataset in order to get a better understanding of the nature of data. 
It is known that Principal Component Analysis (PCA) can be used to reduce the number of features to two principle components and this enables us to visualize the dataset \cite{ng}. 
The first and second principle components resulting from PCA were used to train SVM-RBF to plot the decision boundaries in order to understand the behavior of SVM with the given dataset.

\begin{figure}[t!]
	\centering
	\includegraphics[scale=0.6]
	{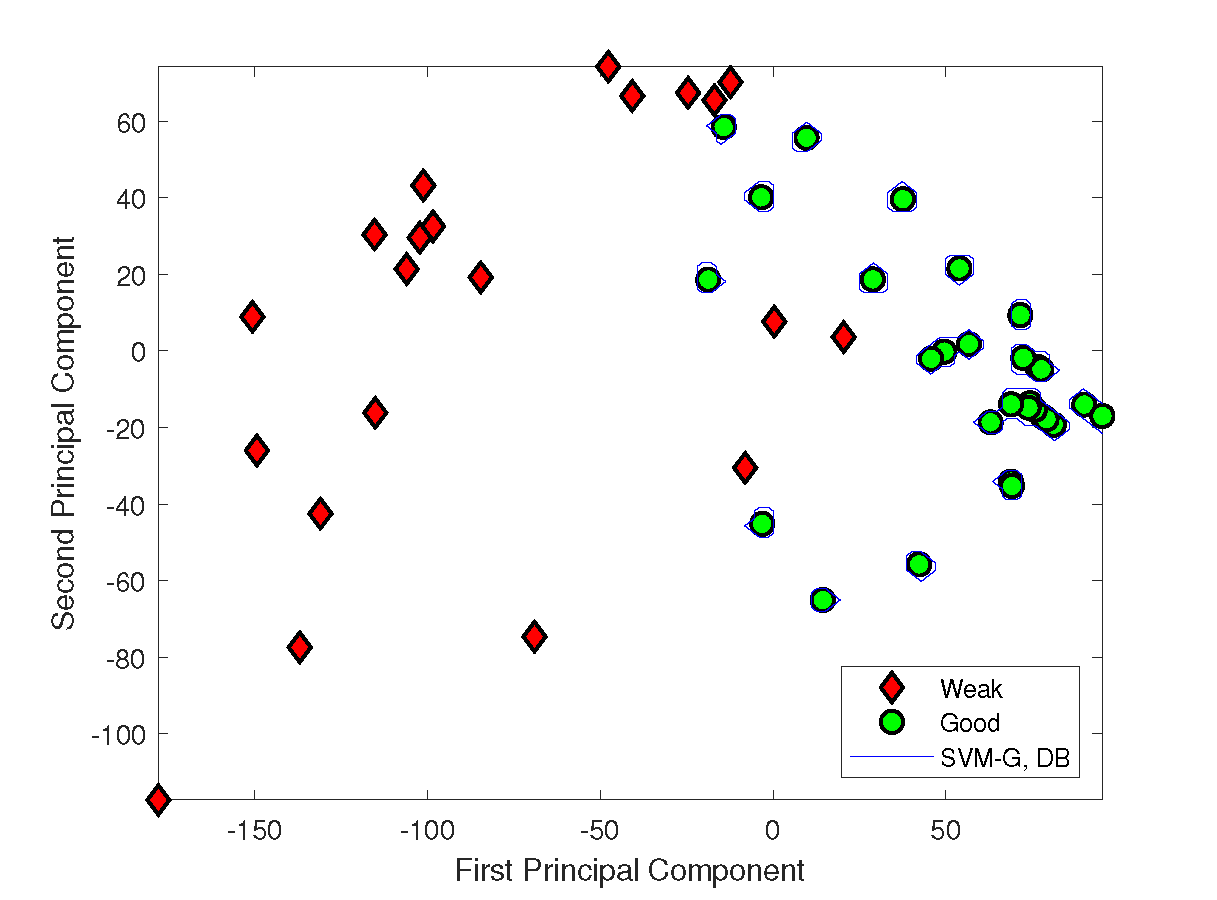}
	\caption{\label{fig:SVM20}Decision boundaries for dataset 1}
\end{figure}

Fig. \ref{fig:SVM20} shows that dataset 1 at 50\% is not linearly separable because there are outlier data points. Indeed, if we were to train a linear classifier, we would likely obtain miss-classified points in the test sample. 
\begin{figure}[t!]
	\centering
	\includegraphics[scale=0.6]
	{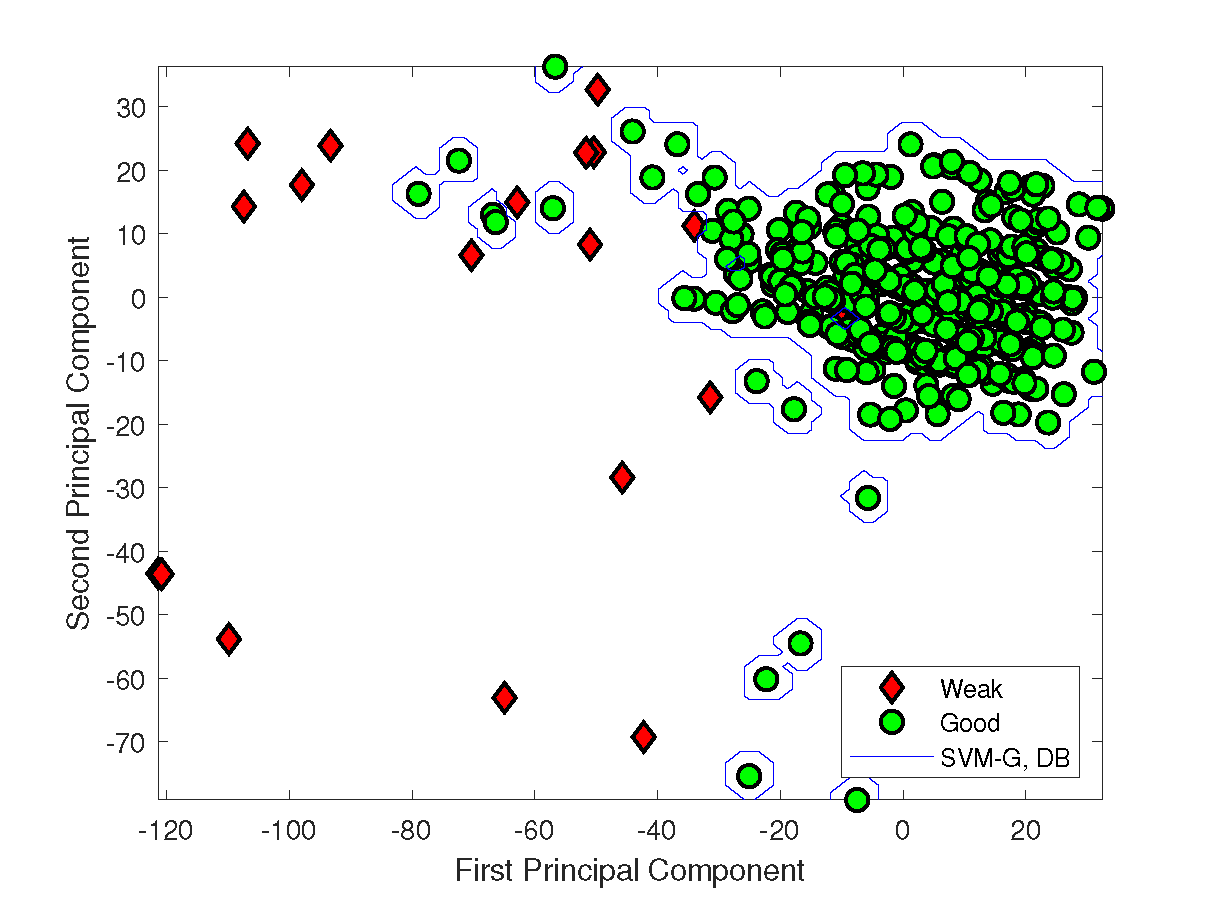}
	\caption{\label{fig:SVM50} Decision boundaries for dataset 2}
\end{figure}

Fig. \ref{fig:SVM50} illustrates the behavior of SVM in building the decision boundary with Gaussian kernel (RBF) of dataset 2 at 50\% stage. In both cases, SVM-RBF model gives a better performance and it is clear that it outperforms the linear kernel (since the data is not linear) and that it is more likely to perform well in classifying new instances.

Moreover, it is shown that PCA shows the overall "shape" of the data \cite{ng}, identifying which samples are similar to one another and which are very different. In other words, PCA can enable us to identify groups of samples that are similar and work out which variables make one group different from another. 

\begin{figure}[h!]
	\centering
	\includegraphics[scale=0.65]
	{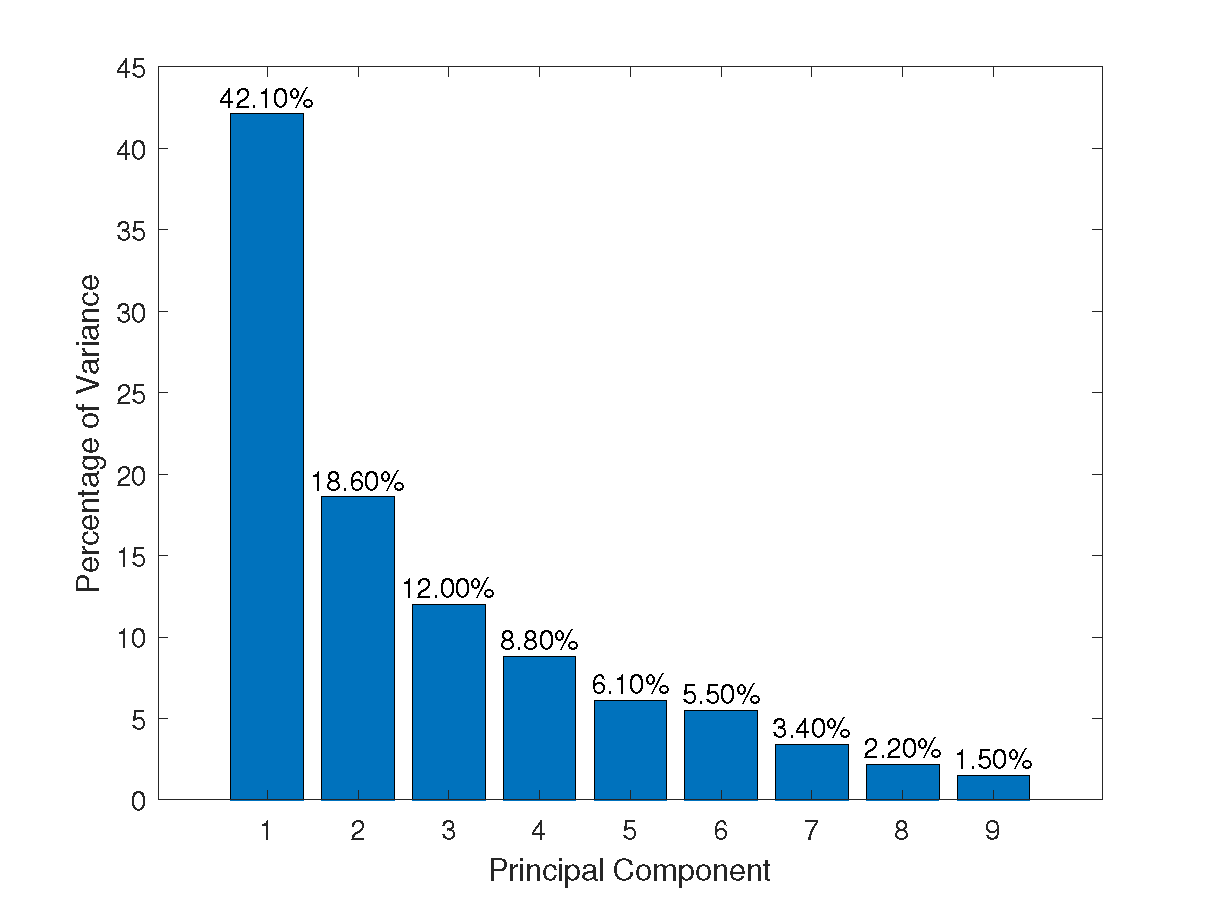}
	\caption{\label{fig:PCA20bars} Dataset 1 - Stage 20\% - Percentage of variance per principal component}
\end{figure}

Performing PCA on Dataset 1 at stage 20\%, we obtain the percentage of variance for every component is as in Figure \ref{fig:PCA20bars}.  Each component explains a percentage of the total variation in the dataset. In particular the first four components can explain 81.5\% of the variance. 

\begin{figure*}[t!]
	\center{\includegraphics[height=8cm,scale=0.6]
		{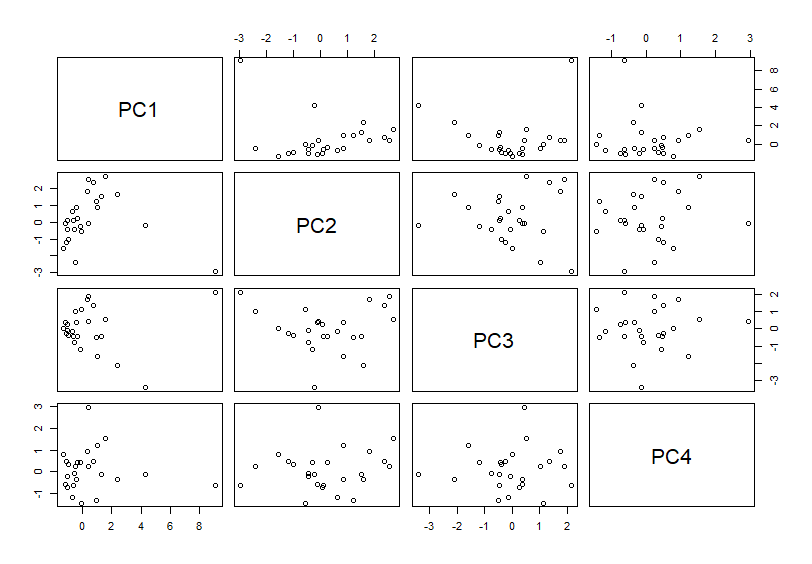}}
	\caption{\label{Dataset1_20_4by4} Dataset 1 - Stage 20\% - First four principal components}
\end{figure*}

\begin{figure*}[t!]
	\center{\includegraphics[scale=0.45]
		{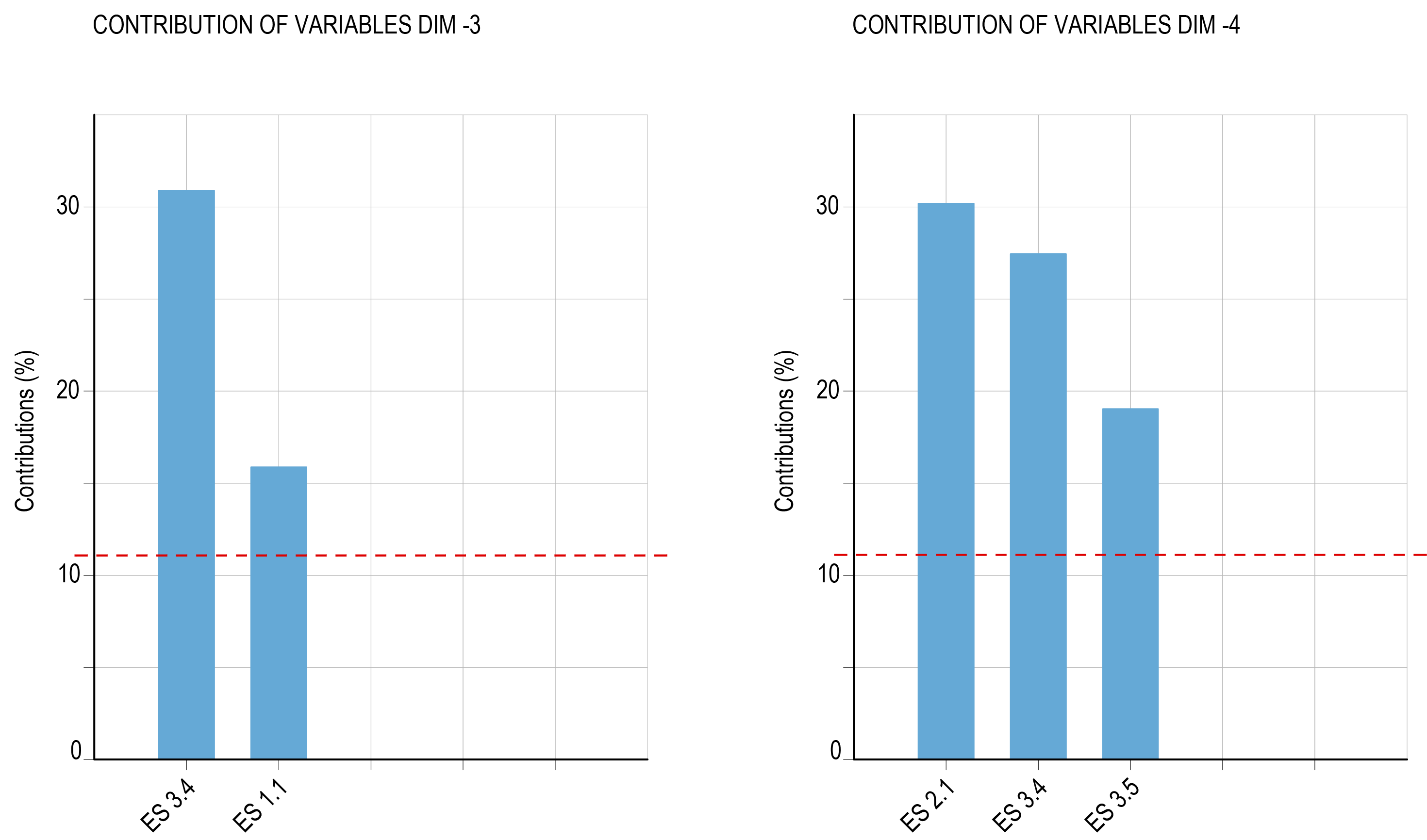}}
	\caption{\label{Dataset1_20_2comp34} Dataset 1 - Stage 20\% - Third and fourth component}
\end{figure*}

For instance, the first principal component PC1 explains 42.1\% of the total variance, which means that almost 1/2 of the information in the dataset can be encapsulated by just that one Principal Component. PC1 and PC2 together can explain 60.7\% of the variance as shown in Figure \ref{fig:PCA20bars}.
More generally, we can plot the first four components 2 by 2 obtaining the following plot that shows in particular that there are many outliers, see Figure \ref{Dataset1_20_4by4}.
We visualize the variable contributions to the principal PC1 - PC4, aiming to give an interpretation of each principal component (see Figures \ref{Dataset1_20_2comp12}, \ref{Dataset1_20_2comp34}):

\begin{figure*}[t!]
	\center{\includegraphics[scale=0.45]
		{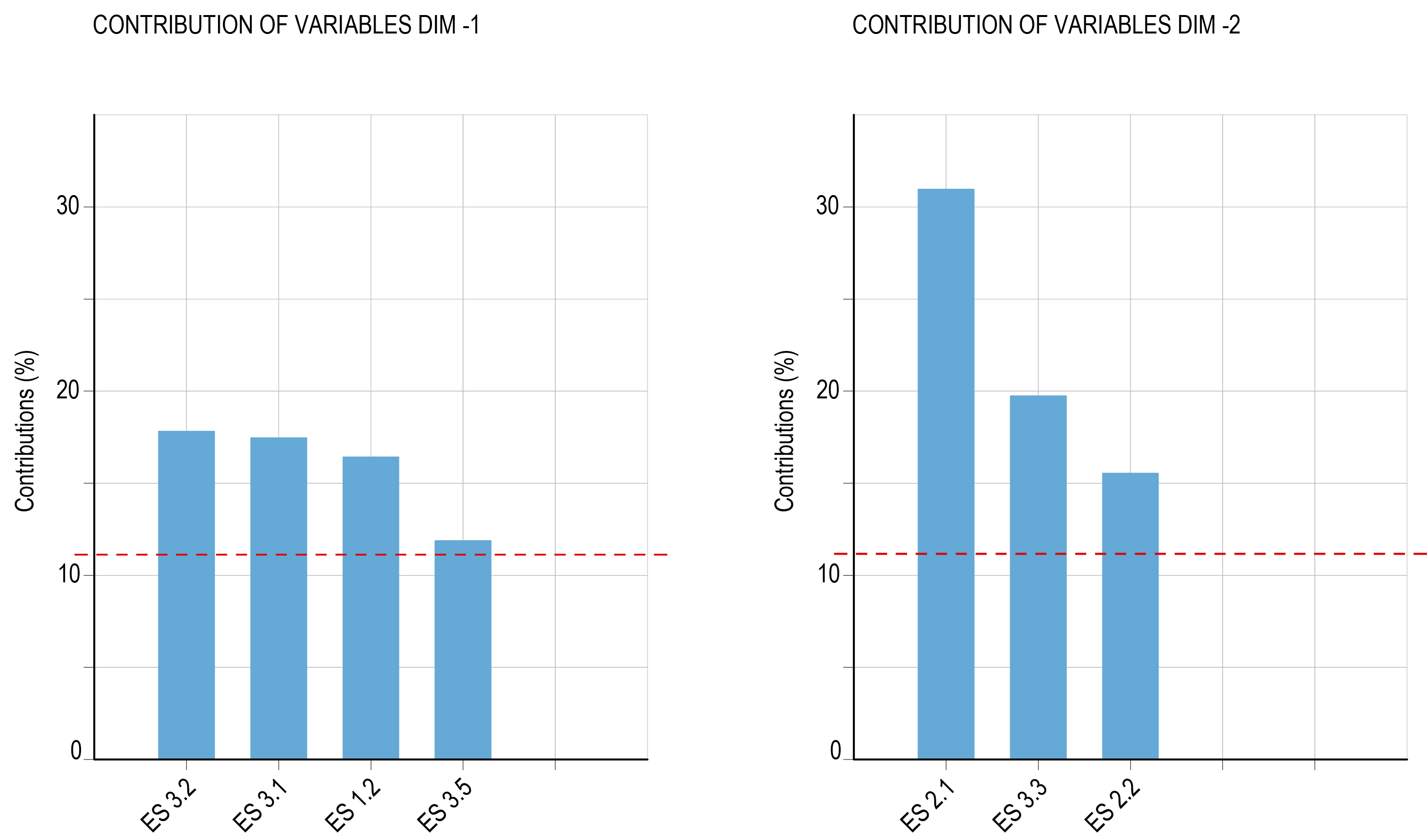}}
	\caption{\label{Dataset1_20_2comp12} Dataset 1 - Stage 20\% - First and second component}
\end{figure*}

So we deduce that:
\begin{itemize}
	\item PC1 corresponds to the \emph{Analyze Task} cluster
	\item PC2 corresponds to the \emph{Apply Task} cluster
	\item PC3 corresponds to the \emph{Understand Task} cluster
	\item PC4 corresponds to the \emph{Evaluate Task} cluster
\end{itemize}
And all these tasks are in Boolean Algebra.

Analogously, we perform PCA on Dataset 1 at stage 50\%, obtaining the percentage of variance for every component. In particular the first four components can explain 76\% of the variance. 
More specifically, the first principal component PC1 in this case explains 40.9\% of the total variance, which means that about 2/5 of the information in the dataset is described by just the first Principal Component.
PC1 and PC2 together can explain 57.8\% of the variance.
\begin{itemize}
	\item PC1 corresponds to the \emph{Evaluate Task} cluster
	\item PC2 corresponds to the \emph{Apply  Task} cluster
	\item PC3 corresponds to the \emph{Analyze  Task} cluster
	\item PC4 corresponds to the \emph{Understand  Task} cluster
\end{itemize}
Accordingly, it can be inferred that tasks that fall under the \emph{Evaluate} and \emph{Analyze} categories based on Bloom's taxonomy (PC1 and PC2 in this case) are better indicators and predictors of student performance. This is because these task categories show the highest level of comprehension of the course material from the educational point of view. Hence, the performance of students in these tasks can provide us with intuitive insights into their overall projected performance in the course. 
\subsection{ML Algorithms' Parameter Tuning }\label{sec:ML_app}
In this section, we describe the classifiers we built for each of the datasets, then we explain the approach used to select the best ensemble learners for the considered datasets. Note that the models were trained on the \emph{raw} datasets and not on the principal components. 
R was used to implement six classifiers and the ensemble learners. The six classifiers that we trained are SVM-RBF, LR, NB, k-NN, RF, and MLP. All the classifiers were trained using all the variables available and maximizing the Gini Index of a 3-fold cross validation \cite{CC1}. Note that a 3-fold cross validation was done in order to reduce the variance. This is based on the fact that using a smaller value of $k$ for cross validation often results in a smaller variance and a higher bias \cite{dd}. On the other hand, 5 different splits of data were used to reduce the bias of the models under consideration.\\ 
The parameters used for each model are tuned using the \emph{grid search} optimization method in such a manner that the Gini Index is maximized. Grid search optimization method is a common method used to hyper tune the parameters of ML classification algorithms. In essence, it discretizes the values for the parameter set \cite{ee}. Models are then trained and assessed for all possible combinations of these values for all the parameters of the ML model used. Despite the fact that this may seem computationally heavy, grid search method benefits from the ability to perform the optimization in parallel, which results in a lower computational complexity \cite{ee}.
 
Table \ref{tab:ML_model_parameter_range} summarizes the range of values for the parameters of the different ML algorithms considered in this work.    
\begin{table}[h!]
	\centering
	\caption{Grid Search Parameter Tuning Range}
	\scalebox{0.9}{
		\begin{tabular}{|p{1.7cm}|p{5.5cm}|p{5.25cm}|} 
			\hline     
			\textbf{Algorithm} & \textbf{Parameter Range in Dataset 1}& \textbf{Parameter Range in Dataset 2}  \\
			\hline
			SVM-RBF & C=[0.25, 0.5, 1] \& sigma = [0.05-0.25] & C=[0.25, 0.5, 1] \& sigma = [0.3-0.8]  \\ \hline
			NB&usekernel=[True,False]&usekernel=[True,False]\\ \hline
			K-NN & k=[5,7,9,...,43]&k=[5,7,9,...,43] \\ \hline
			MLP& number of neurons = [1,3,5] \& number of hidden layers = 1 & number of neurons = [1,3,5] \& number of hidden layers = 1 \\ \hline
			RF & mtry=[2,3,...,12]&mtry=[2,3,4]\\ \hline
	\end{tabular}}
	\label{tab:ML_model_parameter_range}
\end{table}

Note the following:
\begin{itemize}
	\item For the NB algorithm, the \emph{usekernel} parameter represents the choice of the density estimator used. More specifically, \emph{usekernel = true} implies that the data distribution is non-Gaussian and \emph{usekernel=false} implies that the data distribution is Gaussian.
	\item The LR algorithm was not included in the table because it has no parameters to optimize. The default function (namely the sigmoid function) was used by the grid search method to maximize the Gini index.
\end{itemize}

For each algorithm and each dataset we show the list of the features ordered by their importance, i.e. their impact on the predictions. This is meant to give only a rough idea of what the most important features are for each algorithm and each dataset, as the ordering, for such small datasets, heavily depends on the split in Train-Test samples chosen. For this reason, the weights of the predictors will not be specified. 

The final step was to select, for each problem, the best ensemble learner among all the possible ensemble learners that could be produced with the six classifiers. 
\subsubsection{Dataset 1 - Stage 20\%}
\begin{itemize}
	\item \emph{RF:}
	The classifier was trained using k-fold cross-validation with $k=3$. 
	\begin{figure}[t!]
		\centering
		\includegraphics[scale=0.6]
		{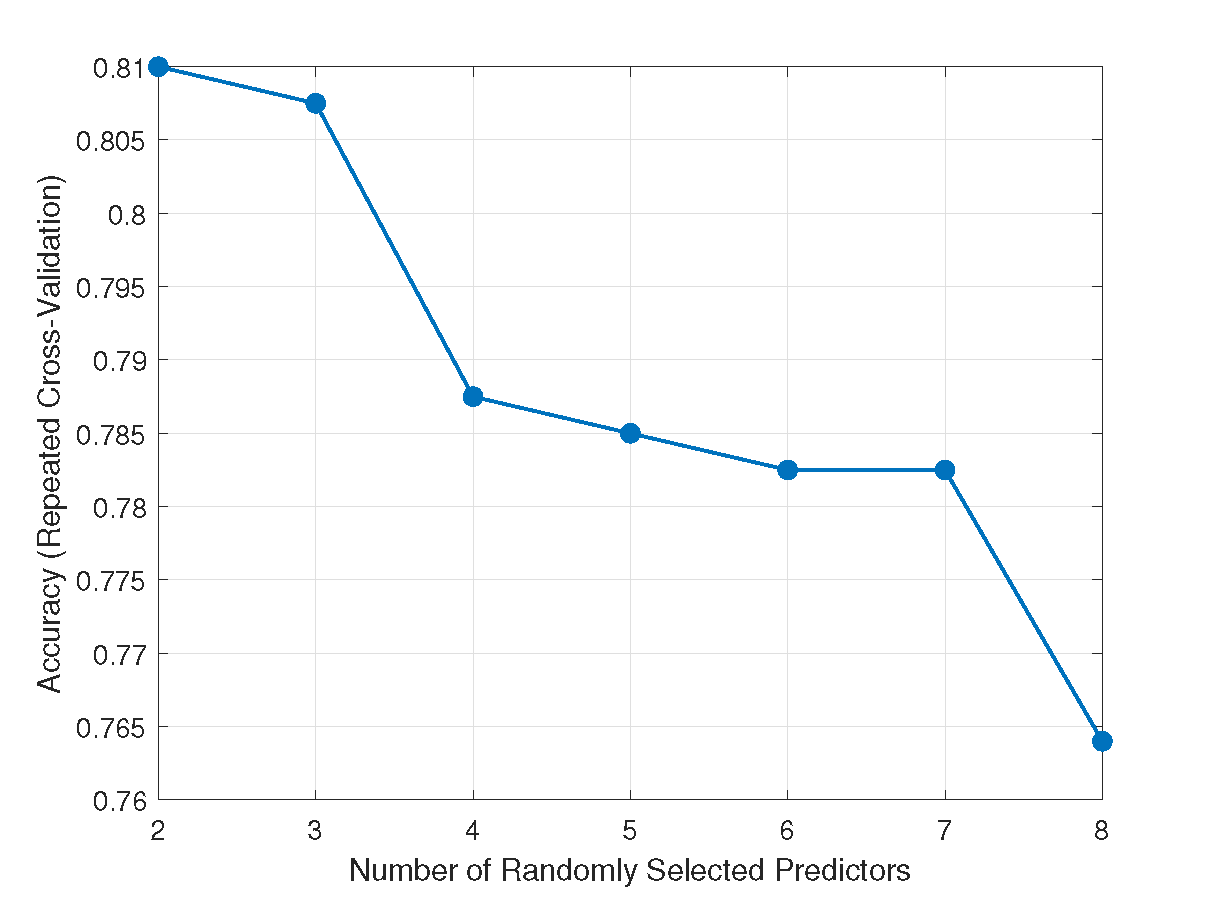}
		\caption{\label{Dataset1_20_RF_vars}Dataset 1 - Stage 20\% - Random Forest, Accuracy vs. mtry value}
	\end{figure}
	Figure \ref{Dataset1_20_RF_vars} shows how the performance changes by choosing a different \emph{mtry} value, i.e. the number of variables available for splitting at each tree node. For example, the optimal value for the \emph{mtry} parameter in the first split was determined to be 3.\\
	The variables' importance in terms of predictivity is described in Table \ref{tab:table_dataset_1_vars_weight} that shows that the most relevant feature is ES3.3, followed by ES3.5, whereas  ES1.1 does not have much impact on the predictions.
	\begin{table}[b!]
		\centering
		
		\caption{Dataset 1 - Stage 20\% - Features' Ranking for Different Base Classifiers}
		\scalebox{1}{
			\begin{tabular}{|c|c|c|c|} 
				\hline     
				\textbf{Ranking} & \textbf{RF} & \pbox{3cm}{\mbox{}\\ \textbf{SVM-RBF, MLP,}\\ \textbf{NB, and k-NN}}& \textbf{LR}\\
				\hline
				1 & ES3.3 &ES2.2 &ES2.2  \\
				\hline
				2 & ES3.5 &ES3.3 &ES3.3\\
				\hline
				3 & ES2.2 &ES3.5 &ES3.5 \\
				\hline
				4 & ES3.4 &ES3.1 &ES3.4\\
				\hline
				5 & ES3.1 &ES3.4 &ES2.1\\
				\hline
				6 & ES1.2 &ES2.1 &ES1.1\\
				\hline
				7 & ES3.2 &ES1.2 &ES3.1 \\
				\hline
				8 & ES2.1 &ES1.1 &ES3.2 \\
				\hline
				9 & ES1.1 &ES3.2 &ES1.2 \\
				\hline
		\end{tabular}}
		\label{tab:table_dataset_1_vars_weight}
	\end{table}
	
	\item \emph{SVM-RBF}:
	SVM algorithm was trained with radial basis function kernel. Table \ref{tab:table_dataset_1_vars_weight} shows the list of the predictors ordered by their impact on the output.

	In particular we can see from the table that the top three variables are ES2.2, ES3.3 and ES3.5 and that ES1.1 and ES3.2 do not have much impact on the predictions. 
	\item \emph{MLP:}
	The variables' importance for MLP classifier is shown in Table \ref{tab:table_dataset_1_vars_weight}.
	\item \emph{NB:}
	The variables' importance for the NB classifier is the same obtained for MLP and SVM as shown in Table \ref{tab:table_dataset_1_vars_weight}.
	\item \emph{k-NN:}
	We trained k-NN classifier trying different values for $k$. 
	For Dataset 1, stage 20\%, the best performance was obtained for $k=9$ and the list of variables ordered by their importance is shown in Table \ref{tab:table_dataset_1_vars_weight}, which is the same as the one obtained for MLP, NB, and SVM classifiers.

	\item \emph{LR:}
	For the LR classifier, variables ES3.1, ES3.2 and ES1.2 have no impact on the predictions. The most important variables for this algorithm are ES2.2 and ES3.3, as shown in Table \ref{tab:table_dataset_1_vars_weight}.
\end{itemize}
In general, the most important features for all the classifiers are ES2.2, ES3.3 and ES3.5, that contributed to the first and second principal components, as we saw in Figure \ref{Dataset1_20_2comp12}.   
\subsubsection{Dataset 1 - Stage 50\%}
\begin{itemize}
	\item \emph{RF:}
	The variables' importance in terms of predictivity is described in Table \ref{tab:table_dataset_1_50_RF_vars_weight} that shows that the most relevant feature is ES4.2, followed by ES4.1, whereas  ES3.2 does not have much impact on the predictions. Also note that the bottom 3 variables are the same as the ones shown for RF classifier on Dataset 1, at stage 20\%.
	\begin{table}[b!]
		\centering
		\caption{Dataset 1 - Stage 50\%  - Features' Ranking for Different Base Classifiers}
		\scalebox{1}{
			\begin{tabular}{|c|c|c|c|} 
				\hline     
				\textbf{Ranking} & \textbf{Feature}& \pbox{3cm}{\mbox{}\\ \textbf{SVM-RBF, MLP,}\\ \textbf{NB, and k-NN}}& \textbf{LR}\\
				\hline
				1 & ES4.2 &ES4.1 &ES4.1 \\
				\hline
				2 & ES4.1 &ES4.2 &ES4.2 \\
				\hline
				3 & ES5.1 &ES5.1 &ES3.3 \\
				\hline
				4 & ES3.5 &ES2.2 &ES1.1 \\
				\hline
				5 & ES3.3 &ES3.3 &ES3.4 \\
				\hline
				6 & ES2.2 &ES3.5 &ES5.1\\
				\hline
				7 & ES3.1 &ES3.1 &ES3.1 \\
				\hline
				8 & ES3.4 &ES3.4 &ES1.2 \\
				\hline
				9 & ES2.1 &ES2.1 &ES2.1 \\
				\hline
				10 & ES1.2 &ES1.2 &ES3.5 \\
				\hline
				11 & ES1.1 &ES1.1 &ES3.2 \\
				\hline
				12 & ES3.2 &ES3.2 &ES2.2 \\
				\hline
		\end{tabular}}
		\label{tab:table_dataset_1_50_RF_vars_weight}
	\end{table}
	
	\item \emph{SVM-RBF:}
	The list of predictors for SVM classifier, ordered by their importance is shown in Table   \ref{tab:table_dataset_1_50_RF_vars_weight}.
	
	\item \emph{MLP:}
	MLP classifier shares with SVM classifier the same table of variables' importance, see Table  \ref{tab:table_dataset_1_50_RF_vars_weight}.
	
	\item \emph{NB:}
	The variables' importance for the NB classifier is the same obtained for MLP, SVM, and k-NN classifiers, see Table \ref{tab:table_dataset_1_50_RF_vars_weight}.
	\item \emph{k-NN:}
	We tried different values for $k$ when we trained k-NN classifier. 
	The final choice of $k$ for dataset 1 at 50\% stage was $k=5$ . As for dataset 1 at 50\% stage , the list of variables ordered by their importance as shown in Table \ref{tab:table_dataset_1_50_RF_vars_weight}, is the same as the one obtained for MLP and SVM classifiers.
	\item \emph{LR:}
	For the LR classifier, variables ES2.2, ES3.2 and ES1.2 have no impact on the predictions. The most important variable for this algorithm is ES4.1 as Table \ref{tab:table_dataset_1_50_RF_vars_weight} shows.
\end{itemize}
In general, the most important features for almost all the classifiers are ES4.1, ES4.2 and ES5.1, that contributed to the first principal component. The only classifier that does not have ES5.1 in the top three variables is the LR classifier which has ES3.3 in third position instead. Also variable ES3.3 belongs to the first principal component. This further validates the previously obtained results during the PCA analysis performed which illustrated the significance and contribution of each of the features in predicting the students' final grade.

\subsubsection{Dataset 2 - Stage 20\%}
We have only two features for Dataset 2, Stage 20\% and for all the classifiers, the list of features ordered by importance is the same, see Table \ref{tab:table_dataset_2_20_predictors}.
\begin{table}[h!]
	\centering
	\caption{Dataset 2 - Stage 20\% - Features' rankings}
	\scalebox{1}{
		\begin{tabular}{|c|c|} 
			\hline     
			\textbf{Ranking} & \textbf{Feature}  \\
			\hline
			1 &Assignment01 \\
			\hline
			2 & Quiz01 \\
			\hline
	\end{tabular}}
	\label{tab:table_dataset_2_20_predictors}
\end{table}

\subsubsection{Dataset 2 - Stage 50\%}
For RF, SVM, MLP, k-NN and NB the lists of features are ordered in the same way while for LR the list order by importance is slightly different, as shown in Table \ref{tab:table_dataset_2_50_predictors}.
\begin{table}[h!]
	\centering
	
	\caption{Dataset 2 - Stage 50\% - Features' Ranking for Different Base Classifiers}
	\scalebox{1}{
		\begin{tabular}{|c|c|c|} 
			\hline     
			\textbf{Ranking} &  \pbox{3cm}{\mbox{}\\ \textbf{RF, SVM,}\\ \textbf{MLP, k-NN and NB}}&\textbf{LR} \\
			\hline
			1 &Assignment02& Assignment02\\
			\hline
			2 & Assignment01&Quiz01\\
			\hline
			3 & Quiz01& Midterm Exam\\
			\hline
			4 & Midterm Exam& Assignment01\\
			\hline
	\end{tabular}}
	\label{tab:table_dataset_2_50_predictors}
\end{table}


Based on the aforementioned results, it  can be seen that assignments are better indicators of the student performance. This can be attributed to the fact that students have more time to complete assignments and are often allowed to discuss issues and problems among themselves. Thus, students not performing well in the assignments can be an indication that they are not fully comprehending the material, resulting in potentially lower overall final course grade.

\subsection{Proposed Ensemble learning model selection: a systematic approach}
For each dataset and at each stage, a systematic approach was used to select the best subset of classifiers to consider for the ensemble learner. More specifically, the procedure was to evaluate the performance of every possible combination of the classifiers that we trained. 

The performance of each model was measured in terms of Gini Index. We inferred each model on the test sample producing a score. Each score corresponds to the probability of being a Weak student.  Note that although confusion matrices present a clear picture of correct and incorrect classifications for each class of objects, they are affected by the choice of a threshold on the probability score. For this reason,  we will rely on the Gini Index instead of evaluating every model using the confusion matrices as it is more robust and less dependent on the probability of the threshold. 

The statistical significance of our results is determined by computing the p-values. The general approach is to test the validity of a claim, called the \emph{null hypothesis}, made about a population. An alternative hypothesis is the one you would believe if the null hypothesis is concluded to be untrue. A small p-value ($\leq 0.05$) indicates strong evidence against the null hypothesis,  so you reject the null hypothesis.
\begin{figure}[t!]
	\centering
	\includegraphics[scale=0.7]{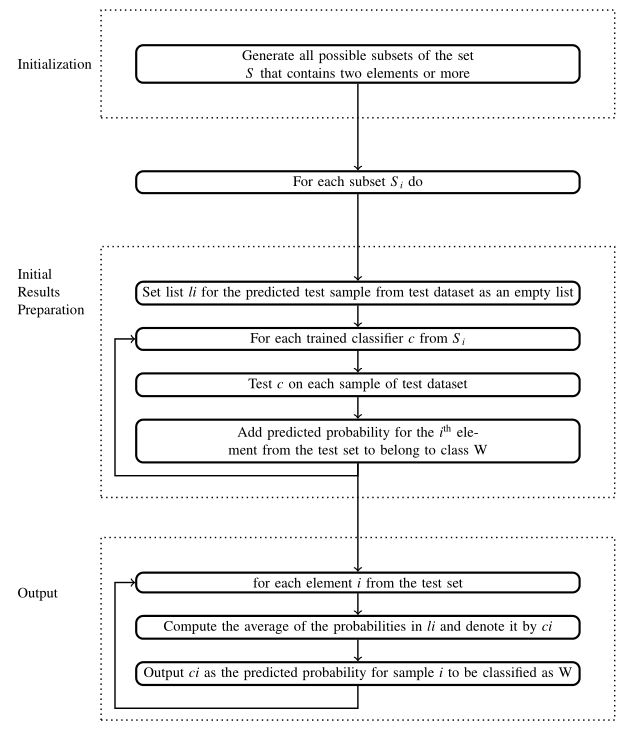}
	\caption{\label{fig:1.3} The procedure of generating ensemble learners and creating the score for each ensemble learner}
\end{figure}

For our purposes, the null hypothesis states that the Gini Indices were obtained by chance. 
We generated 1 million random scores from normal distribution  and calculated the p-value. The ensemble learners selected have p-value $\leq 0.05$, indicating that there is strong evidence against the null hypothesis. Therefore, choosing an ensemble model using a combination of Gini Index and p-value allows us to have a more statistically significant and robust model.

\begin{figure}[t!]
	\centering
	\includegraphics[scale=0.6]
	{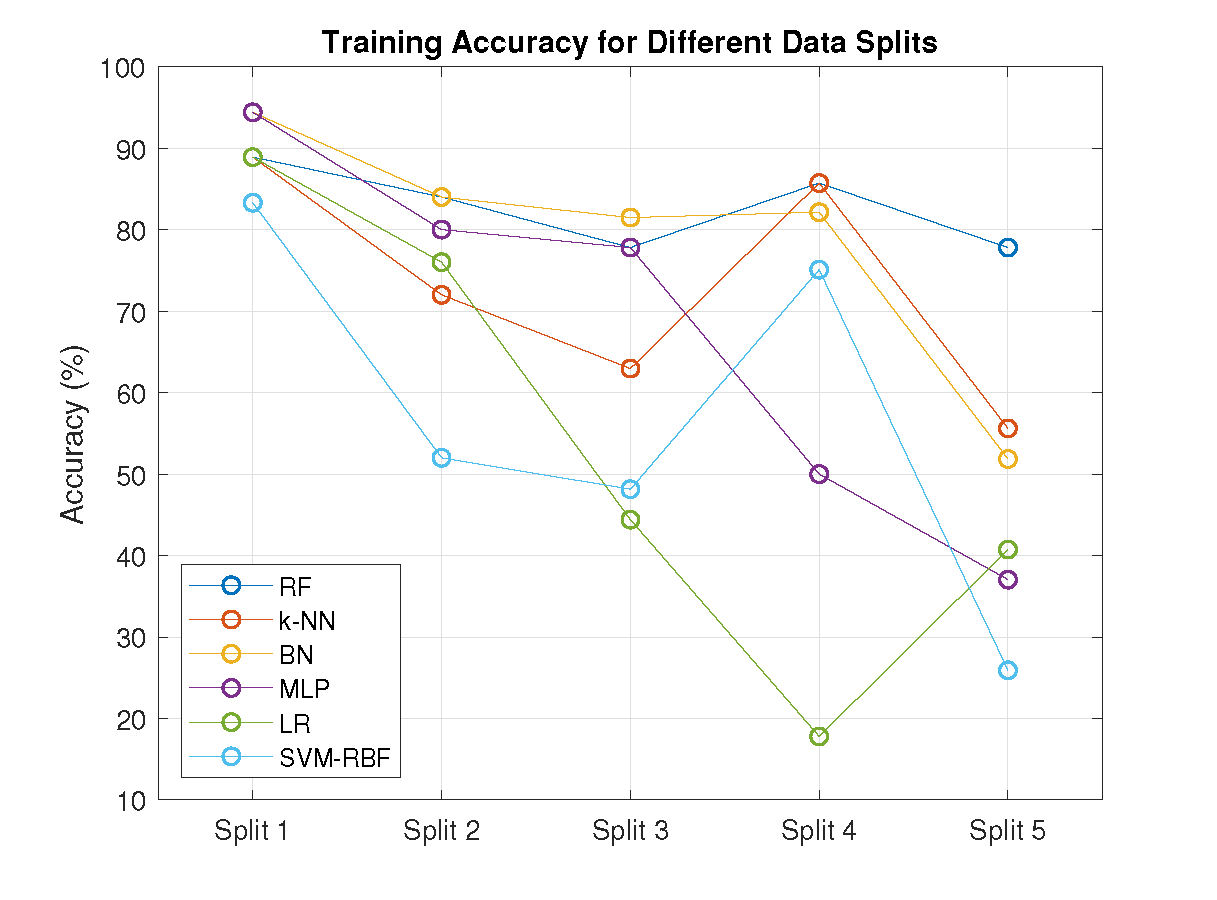}
	\caption{\label{LR_5_splits_small20}{Dataset 1 - Stage 20\% - Performance of different classifiers on 5 splits}}
\end{figure}

For each model we create a matrix consisting of the target variable and the score produced by the model, then we order the matrix by the score in descending order. In this way, on top we can find the students which are more likely to be Weak students as opposed to the bottom of the matrix where we find the students who are less likely to be Weak students. 
Finally, for each dataset, we generate all the possible combinations of the six models and calculate the corresponding Gini Index. 
The procedure followed to produce each ensemble learner can be summarized in Figure \ref{fig:1.3}.
Since the training and test samples are small sized, many of the ensemble learners produced had the same Gini Index and the performances seemed to depend on the split into training and test samples that was chosen at the beginning. For instance, Figure \ref{LR_5_splits_small20} shows the performance of the different classifiers on Dataset 1 at stage 20\% on different splits.  For example, it can be seen that the performance of the LR classifier is really good on the first two splits considered (with Gini Index 88.9\% and 76\% respectively), whereas on the fourth split it performs very poorly (Gini Index is only 17.8\%).
Since the performance of the classifiers depends very much on the split, and so do all the ensembles, instead of considering only one split we considered 5 additional splits of the dataset, namely split1= (Training1, Test1), split2 = ( Training2, Test2), split3 = (Training3, Test3),  split4 = (Training4, Test4),  split5 = (Training5, Test5), and ran the 6 algorithms on each split, training a total of $6 \times 5$ models, each one run using a 3-fold method and keeping track of the performances of each model also on the folds. We average of the performance across the splits to remove any potential bias. Note that every split was created randomly, like for the initial training and test samples, and in such a way that the target variable of each training and test sample was representative of the entire dataset. 

Afterwards, we compare the performances of the models of each split, and produce one table of all possible ensembles.

Although from the literature we expect the ensemble learners to perform better than the individual classifiers, we included in the comparison also the individual classifiers and considered 64 combinations of classifiers as opposed to 57 (the actual ensemble learners).\\
Finally we created a table consisting of 64 rows, each one representing a specific ensemble, where in each row we have:
\begin{itemize}
	\item the first 6 entries, one for each algorithm, with 1's and 0's corresponding to presence or absence of the algorithm in the ensemble. In particular, the individual classifiers correspond to those rows for which the sum of the first six entries is one. 
	\item entries 7,8,9,10,11,12 corresponding to the Gini index, namely G, G1, G2, G3, G4, G5, associated to the initial split, split1, split2, split3, split4, split5.
	\item entry 13, called Avg, corresponding to the average of G, G1, G2, G3, G4 and G5.
\end{itemize}  
with a subset shown in Table \ref{tab:table_dataset_1_best}.

The table was ordered with respect to Avg, in descending order, and the top ensemble was selected as the best classifier. Moreover, the p-values were calculated and shown in the table in the 14-th column called p, and prove that the ensemble selected is statistically significant.

\section{Results and Discussion}\label{sec:disc}
In this section we discuss the results and select an ensemble learner for each of the four experiments. Finally we set a threshold and evaluate the performance based on the corresponding confusion matrices.
\subsection{Results: Dataset 1 - Stage 20\%}
As explained, 30 models were trained, 6 on each of the 5 splits. The top 3 models in terms of Gini Index are RF, NB and k-NN. Once the matrix with all possible ensemble learners is created, we ordered it with respect to Average Gini Index. 

\begin{figure}[t!]
	\center{\includegraphics[scale=0.5]
		{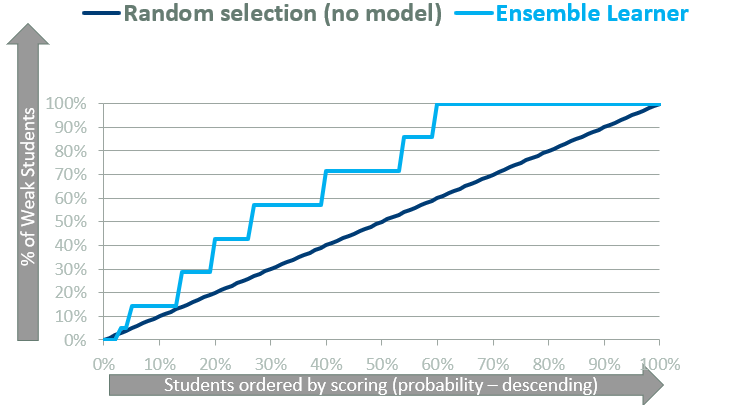}}
	\caption{\label{perf_small20_graph}{Dataset 1 - Stage 20\% - Performance Ensemble Learner}}
\end{figure}

Table \ref{tab:table_dataset_1_best} consists of the best 3 classifiers (one of each row). 
\begin{table*}[b!]
	\centering
	\caption{ Dataset 1 - Stage 20\% - Best classifiers}
	\scalebox{0.75}{%
		\begin{tabular}{|c|c|c|c|c|c|c|c|c|c|c|c|c|c|} 
			\hline
			\textbf{rf}& \textbf{mlp} & \textbf{bn} & \textbf{knn} & \textbf{lreg} & \textbf{svm} & \textbf{G} & \textbf{G1} & \textbf{G2} & \textbf{G3} &\textbf{G4}&\textbf{G5}& \textbf{Avg} & \textbf{p}\\  \hline
			1&0&1&0&0&0&0.750 &0.899&0.880 &0.815 & 0.857 &0.778 & 0.828& 0.0034\\    \hline
			1&0&1&1&0&0&0.679 &0.944&0.840 &0.852 & 0.821 &0.815 & 0.825& 0.0034\\    \hline
			1&0&1&0&0&0&0.786 &0.899&0.840 &0.778 & 0.857 &0.778 & 0.821& 0.0045\\    \hline
	\end{tabular}}
	\label{tab:table_dataset_1_best}
\end{table*}
In the table, the 1's correspond to the presence of the model in the ensemble whereas the 0's indicate that the corresponding model should not be included. 
We select, for Dataset 1 at stage 20\%, the ensemble corresponding to the first row, i.e. the ensemble formed by RF and NB. The Gini Index associated with this ensemble learner for the original split is 75\%.  Although this Gini Index is not the highest reached on the initial dataset, we believe that the ensemble chosen is more robust as it comes from the test on 6 different splits. 
The Gini Index of the ensemble chosen corresponds to the area between the model curve (light blue) and the straight line (in dark blue – no model), in Figure \ref{perf_small20_graph}. 
Furthermore, it was observed that the number of Weak Students decreases by 100 times, moving from Highest scoring to Lowest Scoring.
Although the ensemble we selected does not show either the lowest false positive rate or the highest accuracy, it is more robust than each individual classifier, i.e. depends less on the split, hence it is more reliable when inferred on a new dataset.
\subsection{Results: Dataset 1 - Stage 50\%}

Following the same procedure, we trained 30 models and compared the performances of the inferences on each test. RF and k-NN have the best performance, whereas LR has the worst performance on average on the datasets. The results obtained for the 3-folds agree with the ones obtained on the test samples: the top 3 models in terms of Gini Index are RF, NB and k-NN.

\begin{table*}[b!]
	\centering
	\caption{Dataset 1 - Stage 50\% - Best Classifiers}
	\scalebox{0.85}{%
		\begin{tabular}{|c|c|c|c|c|c|c|c|c|c|c|c|c|c|} 
			\hline
			\textbf{rf}&\textbf{mlp}&\textbf{bn}&\textbf{knn}&\textbf{lreg}&\textbf{svm}&\textbf{G}&\textbf{G1}&\textbf{G2}&\textbf{G3}&\textbf{G4}&\textbf{G5}&\textbf{Avg}&\textbf{p}\\  \hline
			1&0&0&0&0&1&0.929&1&1&1&0.929&1&0.976&0.00036 \\       \hline
			1&0&0&1&0&1&0.929&1&1&1&0.929&1&0.976&0.00036 \\       \hline
			1&1&0&1&0&1&0.929&1&1&1&0.929&1&0.976&0.00036 \\       \hline
	\end{tabular}}
	\label{tab:table_dataset_1_50best}
\end{table*}

The best 3 ensembles are shown in Table \ref{tab:table_dataset_1_50best}. The top three rows of the matrix of all possible ensemble learners, that was ordered with respect to the Avg, have same Gini Index and same p-value. 

\begin{figure}[t!]
	\center{\includegraphics[scale=0.5]
		{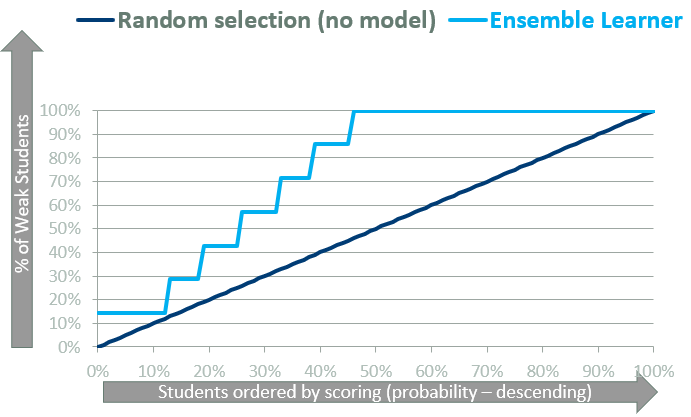}}
	\caption{\label{perf_small50_graph}{Dataset 1 - Stage 50\% - Performance Ensemble Learner}}
\end{figure}

Although they all are good candidates to be selected, we decide not to choose the third ensemble that includes MLP classifier as it performed very poorly on certain splits. Since k-NN had very good performances on all splits, we decide to include it in the ensemble. Accordingly, despite the fact that it may be more computationally expensive, we choose the ensemble corresponding to the second row of Table \ref{tab:table_dataset_1_50best}, i.e. the ensemble formed by RF, k-NN and SVM classifiers to improve the probability of correctly classifying instances based on classifiers' votes.
The ensemble learner has Gini Index = 92.9\%, represented by the area between the model curve and the straight line in Figure \ref{perf_small50_graph}.
In particular, we can see from Figure \ref{perf_small50_graph} that if we order the students by their probability of being a Weak student, we get 60\% of Weak students in the first 30\% of students, and 100\% of Weak students in the first 50\%, as opposed to only 30\% and 50\% respectively if we were not to use the model. 
Similarly, it was observed that the number of Weak Students decreases by 100 times, moving from Highest scoring to Lowest Scoring.

\subsection{Results: Dataset 2 - Stage 20\%}
\begin{figure}[t!]
	\center{\includegraphics[scale=0.5]
		{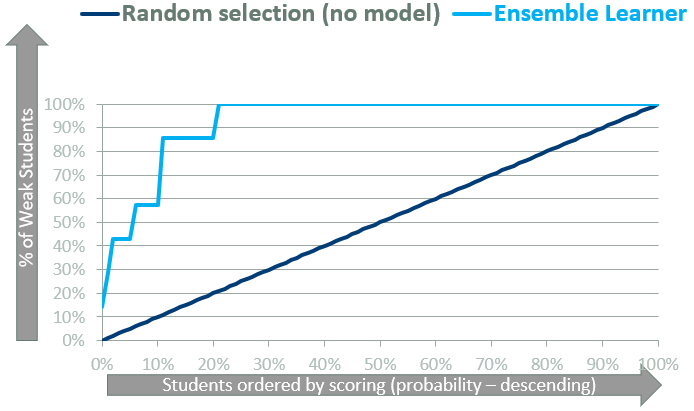}}
	\caption{\label{perf_big20_graph}{Dataset 2 - Stage 20\% - Performance Ensemble Learner}}
\end{figure}

In the same way, we trained 30 models and compared the performances of the inferences on each test. For this dataset, RF, SVM and k-NN do not have good performance. The best classifiers in this case are LR, MLP and NB, and the results obtained for the 3-folds agree with the ones obtained on the test samples. The best 3 ensemble learners for Dataset 2 at stage 20\% are shown in Table \ref{tab:table_dataset_2_20best}.
\begin{table*}[h!]
	\centering
	\centering
	\caption{Dataset 2 - Stage 20\% - Best Classifiers}
	\scalebox{0.7}{%
		\begin{tabular}{|c|c|c|c|c|c|c|c|c|c|c|c|c|c|} 
			\hline
			\textbf{rf}&\textbf{mlp}&\textbf{bn}&\textbf{knn}&\textbf{lreg}&\textbf{svm}&\textbf{G}&\textbf{G1}&\textbf{G2}&\textbf{G3}&\textbf{G4}&\textbf{G5}&\textbf{Avg}&\textbf{p}\\  \hline
			0&0&1&0&1&0&0.89&0.698&0.872&0.846&0.849&0.863&0.8363&0.0000024 \\       \hline
			0&0&0&0&1&0&0.888&0.702&0.876&0.84&0.856&0.854&0.8360&0.0000024 \\       \hline
			0&1&1&0&0&0&0.882&0.667&0.872&0.834&0.867&0.894&0.8360&0.0000032 \\       \hline
	\end{tabular}}
	\label{tab:table_dataset_2_20best}
\end{table*}
Hence, for this dataset we select the ensemble learner formed by NB and LR. The Gini Index of the ensemble selected is 89\% and is represented by the area between the model curve and the straight line as per Figure \ref{perf_big20_graph}.
In particular, 100\% of Weak students are identified by the ensemble learner in the first 21.4\% of students ordered by scoring, in descending order. 

\subsection{Results: Dataset 2 - Stage 50\%}
For this dataset, RF, SVM and k-NN do not have good performance. The best classifiers in this case are LR, MLP, followed by NB. 
For Dataset 2 at stage 50\% we select the ensemble learner formed by MLP and LR as shown in Table \ref{tab:ensembles_big20part22}. 
\begin{table*}[h!]
	\centering
	\caption{Dataset 2 - Stage 50\% - Best Classifiers}
	\scalebox{0.7}{%
		\begin{tabular}{|c|c|c|c|c|c|c|c|c|c|c|c|c|c|} 
			\hline
			\textbf{rf}&\textbf{mlp}&\textbf{bn}&\textbf{knn}&\textbf{lreg}&\textbf{svm}&\textbf{G}&\textbf{G1}&\textbf{G2}&\textbf{G3}&\textbf{G4}&\textbf{G5}&\textbf{Avg}&\textbf{p}\\  \hline
			0&1&0&0&1&0&0.899&0.888&0.925&0.977&0.929&0.988&0.934&0.00012 \\       \hline
			0&0&0&0&1&0&0.934&0.876&0.923&0.98&0.902&0.983&0.933&0.00001 \\       \hline
			0&1&0&0&0&0&0.859&0.886&0.932&0.98&0.929&0.981&0.928&0.00075 \\       \hline
	\end{tabular}}
	\label{tab:ensembles_big20part22}
\end{table*}
\begin{figure}[h!]
	\center{\includegraphics[scale=0.45]
		{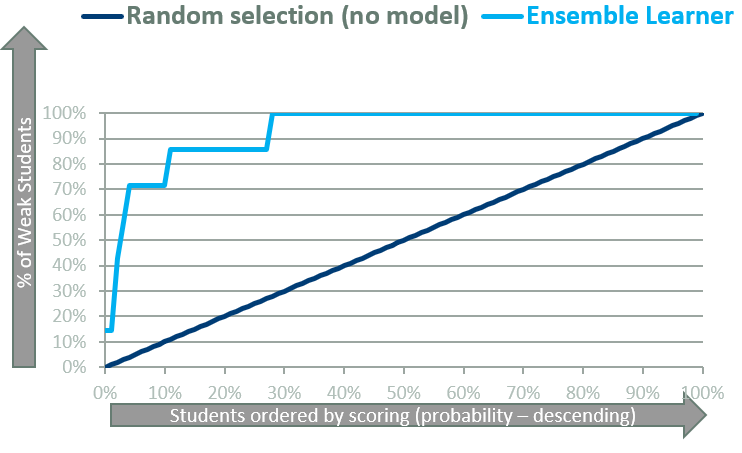}}
	\caption{\label{perf_big50_graph}{Dataset 2 - Stage 50\% - Performance Ensemble Learner}}
\end{figure}
The Gini Index of the ensemble selected is 89.9\% and is represented by the area between the model curve and the straight line, see Figure \ref{perf_big50_graph}. In particular, 100\% of Weak students are identified by the ensemble learner in the first 28.28\% of students ordered by their probability of being a Weak student.\\ 
\indent Table \ref{perf_mat_base} summarize the specificity and sensitivity results of all the base learners for the two datasets. It is worth noting that the performance was evaluated using the specificity and sensitivity due to the fact that the datasets studied are imbalanced. This is a regular occurrence in educational datasets.
\begin{table*}[h!]
	\centering
	\caption{Base Classifiers' Performances\label{perf_mat_base}}
	\scalebox{0.7}{%
		\begin{tabular}{|l|c|c|c|c|}
		\hline
		\multicolumn{5}{|c|}{\textbf{Specificity}}\\  \hline
		\textbf{Technique}&\multicolumn{2}{c|}{\textbf{Dataset 1}}&\multicolumn{2}{c|}{\textbf{Dataset 2}}\\ \hline
		&Stage 20\%&Stage 50\%&Stage 20\%&Stage 50\%\\ \hline
		RF&0.75&0.875&0.891&0.956\\ \hline
		MLP&0.875&0.75&0.949&0.992\\ \hline
		K-NN&0.75&0.875&0.949&0.992\\ \hline
		NB&0.875&1&0.942&0.963\\ \hline
		LR&0.75&0.75&0.949&0.992\\ \hline
		SVM&0.75&0.75&0.949&0.956\\ \hline
		\multicolumn{5}{|c|}{\textbf{Sensitivity}}\\  \hline
		RF&1&0.857&0.857&0.714\\ \hline
		MLP&0.714&1&0.714&0.714\\ \hline
		K-NN&0.857&0.857&0.571&0.428\\ \hline
		NB&0.428&0.857&0.714&0.857\\ \hline
		LR&0.857&0.857&0.714&0.714\\ \hline
		SVM&0.714&0.857&0.571&0.714\\ \hline
		\end{tabular}}
\end{table*}

\subsection{Ensemble Learners}
The ensemble learners selected are formed by:
\begin{enumerate}
	\item RF and NB for Dataset 1 at stage 20\%
	\item RF, k-NN and SVM for Dataset 1 at stage 50\%
	\item NB and LR for Dataset 2 at stage 20\%
	\item MLP and LR for Dataset 2 at stage 50\%
\end{enumerate}
It is worth noting that the RF classifier was chosen as part of the ensemble model in both stages for dataset 1. This is mainly due to the feature-rich nature of this dataset despite the low number of instances available. On the other hand, it can be seen that LR classifier was an integral part of the ensemble model for dataset 2. This can be attributed to two main factors. The first is that LR requires a relatively large sample size which is the case for dataset 2, but not dataset 1. The second is that LR classifier assumes that features are independent of each other. In case of dataset 2, due to the low number of features considered at the 20\% and 50\% stages, the correlation between the features may not be as evident, which can result in the LR classifier assuming them to be independent. Hence, the LR classifier performed well as an individual classifier at both stages and accordingly was included as part of the ensemble model. 

Table \ref{perf_mat_ensembles} illustrates the performances of the ensemble learners in terms of accuracy, precision, recall/sensitivity, F-measure, and specificity. These quantities depend on the \emph{threshold} $\tau$.
\begin{table*}[h!]
	\centering
	\caption{Ensemble Learners' Performances\label{perf_mat_ensembles}}
	\scalebox{0.8}{%
		\begin{tabular}{l|c||c|c|c|c|c} 
			&                                              $\tau$        & \textbf{Accuracy} & \textbf{Precision} & \textbf{Sensitivity} & \textbf{F-Measure} & \textbf{Specificity}\\
			\hline
			\text{Dataset 1 - stage 20\%}&   0.35              & 0.800 & 0.833& 0.714 & 0.769 & 0.875 \\
			\text{Dataset 1 - stage 50\%} &   0.35             &  0.867 & 0.857 & 0.857 & 0.857 & 0.875 \\
			\text{Dataset 2 - stage 20\%}&    0.065           & 0.966 & 0.625 & 0.714 & 0.667 & 0.978 \\
			\text{Dataset 2 - stage 50\%} &  0.2                &  0.917 & 0.353 & 0.857 & 0.500 & 0.920 \\
	\end{tabular}}
\end{table*}

Based on the results shown in Table \ref{perf_mat_ensembles}, it can be seen that the proposed ensemble models achieve high accuracy and high specificity. This means that the proposed ensemble model selection approach resulted in providing a model that can help instructors identify students who may need help during the course delivery. In turn, this would allow the instructors to have a more proactive role in helping their students. As mentioned earlier, the performance was evaluated using the specificity and sensitivity due to the fact that the datasets studied are imbalanced. This is a regular occurrence in educational datasets.

\section{Research Limitations}\label{sec:limitation}
Despite the promising results obtained using the proposed approach, this work suffers from some limitations that may have affected the results. 
\begin{itemize}
	\item[A] For Dataset 1, the main issue was the size: only 52 students could be considered for our experiment, and the models were trained on only 70\% of them and tested on the remaining 30\%, corresponding to a number of students which is not statistically relevant.\\
	 Although it is possible to use over-sampling techniques such as Synthetic Minority Over-sampling Technique (SMOTE) to increase the dataset sample size, such techniques increase the complexity of the model and may lead to model over-fitting \cite{ch4_rev1d}. Thus, using such techniques was not considered in this work in order to maintain the real-life nature of the dataset under consideration.
	\item[B ]For Dataset 2, at the 20\% stage, the number of student was not an issue, but we could only use two features to build the classifiers.
\end{itemize}
In both cases it was not possible to obtain additional data: indeed, for the second dataset it took almost a year to get the data because of the privacy.
\begin{itemize}
	\item[C] Another main factor is that there are many outliers, i.e. points that have very different characteristics from all the other points of the dataset as seen in Figures \ref{fig:SVM50} and \ref{fig:SVM20}. These points correspond to those students who had a good performance at all tasks except for one, where they did not perform it, getting zero grade (e.g. getting a zero grade in the midterm). The classifiers are more likely to give a wrong prediction for these students. However, these outlier points could not be removed as this would threaten the integrity and validity of the proposed analysis. 
	\item[D] Another issue encountered was that Dataset 2 is unbalanced, i.e. the percentage of weak students in the target variable is very low.
	\item[E] As we have seen in Section \ref{Dataset_Visualization}, the datasets are non-linear and consequently any linear classifier would not perform well on such given datasets. 
\end{itemize}
It is worth mentioning that these challenges and limitations are common when dealing with educational datasets. However, despite all the issues encountered, we highlight that \emph{the classifier was able to predict correctly the weak students}, as it was shown in Section \ref{sec:disc}.
\section{Conclusion and Future Work} \label{sec:future}
Educational data mining has garnered significant interest in recent years in an attempt to personalize and improve the learning process for students. Therefore, many researchers have focused on trying to predict the performance of learners. However, such a task is not simple to achieve, especially during the course delivery. To that end, this work thoroughly explored and analyzed two different datasets at two separate stages of course delivery (20\% and 50\% respectively) using multiple graphical, statistical, and quantitative techniques. This analysis showed the non-linear nature of the collected data in addition to the correlation between the features. These insights were then used to help choose and tune the classification algorithms and their parameters respectively. Furthermore, a systematic ensemble learning model selection approach was proposed based on the combination of Gini Index and statistical significance indicator (p-value) to predict students who may need help in an e-learning environment. Experimental results showed that the proposed ensemble models achieve high accuracy and low false positive rate at all stages for both datasets.

Based on the aforementioned research limitations, below are some suggestions for our future work. For instance, the best way to face [A] and [B] would be to have more data available, by collecting training and testing datasets for every time the course is offered. Unfortunately, not all data that would need to be added to our dataset can be collected for privacy reasons. Even though we cannot collect data such as students' personal information, we can collect data regarding their attendance and we believe this feature would further help the classifier be more accurate. 

Issue [C] suggests that we could build a predictive model that aims to classify the outliers: this would be incredibly useful as it would allow us to detect those students who seem to be performing well at first, but that are likely to end up \emph{becoming weak students} because of just one task. We could use this predictive model to try to prevent this from happening. To build such a model, we would need information about the students' attendance. 

A possible solution for issue [D],  is usually given by the \emph{Synthetic Minority Over-sampling TEchnique (SMOTE algorithm)}: this algorithm consists of a combination of under-sampling the majority class (Good students) and over-sampling the minority class (Weak students) by adding Synthetic points to the dataset, \cite{R}. There are other methods to solve problems related to unbalanced datasets, for example one could use k-NN and define outliers to be those points that are the furthest from all the other points, or could use SVM to find a hyperplane that isolates the outliers from the rest of the points. There are many methods and algorithms available in literature, see \cite{T}, and it would be interesting to run experiments using several techniques and compare their performance on this specific dataset.


%

%
%
%
%
%





\bibliographystyle{IEEEtran}
\bibliography{ref}

\balance

\end{document}